\documentclass[aip,jcp,reprint,superscriptaddress,final]{revtex4-1}
\usepackage{graphicx}
\usepackage{dcolumn}
\usepackage{bm}

\usepackage{showkeys}
\usepackage[usenames,dvipsnames]{color} 
\usepackage{mathrsfs}
\usepackage{amsmath}
\usepackage{amssymb}

\newcommand{\beq}{\begin{eqnarray}}
\newcommand{\eeq}{\end{eqnarray}}
\newcommand{\dd}{{ \! \! \rm d}}
\newcommand{\rr}{{\bf r}}
\newcommand{\RR}{{\bf R}}

\begin{document}

\title{Mesoscopic analysis of Gibbs' criterion for sessile nanodroplets \\ on trapezoidal substrates}
  \author{F. Dutka}
    \affiliation{Max Planck Institute for Intelligent Systems, Heisenbergstrasse 3, 70569 Stuttgart, Germany}
    \affiliation{Institute for Theoretical and Applied Physics, University of Stuttgart, Pfaffenwaldring 57, 70569 Stuttgart, Germany}
  \author{M. Napi\'orkowski}
    \affiliation{Institute of Theoretical Physics, University of Warsaw, Ho\.za 69, 00-681 Warszawa, Poland}
  \author{S. Dietrich} 
    \affiliation{Max Planck Institute for Intelligent Systems, Heisenbergstrasse 3, 70569 Stuttgart, Germany}
    \affiliation{Institute for Theoretical and Applied Physics, University of Stuttgart, Pfaffenwaldring 57, 70569 Stuttgart, Germany}

\date{\today}

\begin{abstract}
By taking into account precursor films accompanying nanodroplets on trapezoidal substrates we show that on a mesoscopic level of description one does not observe the phenomenon of liquid-gas-substrate contact line pinning at substrate edges. This phenomenon is present in a macroscopic description and leads to non-unique contact angles which can take values within a range determined by the so-called Gibbs' criterion. Upon increasing the volume of the nanodroplet the apparent contact angle evaluated within the mesoscopic approach changes continuously between two limiting values fulfilling Gibbs' criterion while the contact line moves smoothly across the edge of the trapezoidal substrate. The spatial extent of the range of positions of the contact line, corresponding to the variations of the contact angle between the values given by Gibbs' criterion, is of the order of ten fluid particle diameters. 

\end{abstract}


\maketitle

\section{Introduction}

\begin{figure}[htb] 
 \begin{center}
  \includegraphics{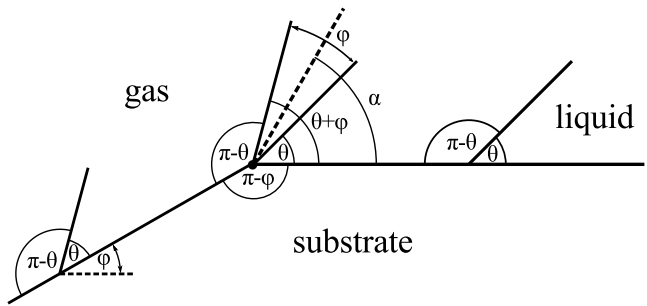}
 \caption{Illustration of Gibbs' criterion for a liquid ridge which is translationally invariant in the direction normal to the plane of the cut shown here. The substrate forms an edge $(\bullet)$ such that its surfaces meet at an angle $\pi-\varphi$. Far from the edge and in thermal equilibrium the liquid wedge forms Young's contact angle $\theta$ with the local substrate surface. If the liquid wedge on the right is pushed left it maintains its contact angle $\theta$ until its three-phase contact line coincides with the edge of the substrate. If it is pushed further, the three-phase contact line remains pinned and the liquid wedge increases its angle with the horizontal substrate surface to a value $\alpha > \theta$ until this angle reaches the value $\alpha=\theta +\varphi$, which coincides with the case of the liquid wedge being far to the left from the edge of the substrate. The angle $\theta+\varphi$ with the horizontal corresponds to Young's angle relative to the tilted substrate surface on the left. Accordingly, if the liquid wedge is pushed further to the left it slides down the tilted substrate surface keeping its local contact angle $\theta$. The same conclusions are reached if the liquid wedge on the far left recedes to the right passing the edge of the substrate. The ambiguity of the local contact angle $\alpha$ if the three-phase contact line coincides with the edge of substrate vanishes in the limit of a planar surface $\varphi \to 0$. This macroscopic picture assumes that (in this cut) all interfaces are straight, geometric lines.  
\label{fig_gibbs}}
 \end{center}
\end{figure}

Recent progress in device miniaturization has led to an increased interest in  adsorption of liquids on substrates structured topographically on the micron- and nanoscale \cite{Herminghaus2008,Rauscher2008,Quere2008,Seemann2005,Seemann2011}. The influence of the  substrate structure on the morphology and on the location of interfaces and three-phase contact lines present in such systems is of particular interest. Already a hundred years ago Gibbs pointed out that an apex-shaped substrate can pin the solid-liquid-gas contact line 
\cite{Gibbs1906}. If a sessile droplet of fixed volume is placed on a planar substrate it forms a spherical cap with a contact angle given by the  modified Young's equation \cite{Swain1998}
\beq \label{mYoung}
 \cos \theta &=& \frac{\gamma_{sg}-\gamma_{sl}}{\gamma} - \frac{\tau}{\gamma} \frac{1}{R} \, ,
\eeq
where $\gamma_{sg}$, $\gamma_{sl}$, and $\gamma$ denote the substrate-gas, substrate-liquid, and liquid-gas surface tension coefficients, respectively; $\tau$ is the line tension coefficient connected with the occurrence of the circular three-phase solid-liquid-gas contact line of radius $R$. For macroscopic droplets or if the droplet is invariant in one spatial direction and forms a ridge, one has $R \to \infty$ and the modified Young's equation reduces to the original Young's equation \cite{Rowlinson1989,Gennes2004}. For a detailed account of the subtleties associated with the line tension see Refs.\,\onlinecite{Schimmele2007,Schimmele2009}.
If the substrate surface forms a sharp corner and the three-phase solid-liquid-gas contact line is located at its apex the modified Young's equation is no longer valid. The corresponding local contact angle  $\alpha$ can take any value within the range \cite{Gibbs1906,Oliver1977,Quere2005,Rauscher2008,Herminghaus2008,Langbein2002} 
\beq \label{gibbs_criterion}
 \theta \leqslant \alpha \leqslant \theta+\varphi \, ,
\eeq
where $\pi-\varphi$ is the angle of the apex formed by the substrate faces (see Fig.\,\ref{fig_gibbs}). This ambiguity of the local contact angle at the apex is called Gibbs' condition or Gibbs' criterion \cite{Kusumaatmaja2008,Blow2009,Toth2011}.
  
For liquid droplets deposited on conical \cite{Chang2010,Extrand2005,Extrand2008} or cylindrical \cite{Du2010,Mayama2011,Wong2009,Toth2011} pillars the solid-liquid-gas contact line can remain pinned at the corresponding sharp edges of the substrates for a range of volumes of the droplets  provided the contact angle fulfills Gibbs' criterion. This fact is widely exploited in the so-called Vapor-Liquid-Solid growth process of nanowires made of semiconductors such as silicon (Si) or germanium (Ge) \cite{Wagner1964,Schmidt2005,Li2007,Roper2010,Ross2005,Schwarz2009,Oh2010,Algra2011,Dubrovskii2011}. In this process a metal sessile droplet is deposited on a substrate exposed to the vapor phase of silicon or germanium. The semiconductor atoms are absorbed by the metal droplet which becomes  supersaturated by them. The ensuing excess semiconductor material precipitates at the boundary of the metal droplet with the substrate, activating the growth of a semiconducting nanowire. Quite often gold droplets are used as a catalyst.

Three-phase contact line pinning and Gibbs' criterion are also crucial for capillary filling in microchannels patterned by posts \cite{Kusumaatmaja2008,Blow2009,Mognetti2010,Chibbaro2009,Mognetti2009,Berthier2009} and for dewetting phenomena on geometrically corrugated  substrates \cite{Ondarcuhu2005}. In the former case, depending on the shape and the height of the posts, and on Young's contact angle $\theta$, the liquid front can be pinned by the posts so that capillary filling of the microchannel might stop. In the dewetting case, the morphology of the emerging holes is modified by the height and the structure of the steps, also due to three-phase contact line pinning.

In addition to the surface and line tension coefficients present in Eq.\,({\ref{mYoung}}), the mesoscopic description of sessile nanodroplets takes into account the effective interface potential acting between the substrate-liquid and the liquid-gas interface. In thermal equilibrium a  nanodroplet with a contact angle less than $180^{\circ}$ is connected with the wetting layer of the liquid phase adsorbed at the substrate \cite{Brochard1991,Gennes1985,Dietrich1988,Bonn2009}. The shape and stability of such effectively two-dimensional ridges or three-dimensional droplets have been discussed in the literature \cite{Yeh1999a,Yeh1999b,Bertozzi2001,Gomba2009,Weijs2011,Doerfler2011,Mechkov2007}. Also the dynamics of nanodroplets was examined on substrates structured geometrically by rectangular steps \cite{Moosavi2006,Moosavi2009}. 

These kinds of mesoscopic studies for planar substrates have not yet been extended to the aforementioned apex-shaped substrate with an arbitrary angle. For such systems, here we focus on how the contact angle changes when the three-phase contact line crosses the edge of the substrate, and whether on the mesoscale the three-phase contact line remains pinned to the edge, as it is the case in the macroscopic description (see Fig.\,\ref{fig_gibbs}).

In Sec.\,II we describe the density functional based effective interface Hamiltonian which enables us to calculate the equilibrium shapes of the liquid-gas interface in the presence of geometrically structured substrates. In Sec.\,III we analyze how the contact angle of the liquid-gas interface separating the coexisting liquid and gas phases varies upon moving across an apex-shaped substrate. It turns out that on the mesoscale the three-phase contact line is not pinned to the edge of the substrate and the contact angle varies in agreement with Gibbs' criterion. The shape of a liquid nanodroplet deposited on a trapezoidal substrate is examined in Sec.\,IV. Upon increasing the volume of this nanodroplet the three-phase contact line  moves smoothly across the edge of the trapezoidal substrate while the apparent contact angle changes continuously between two limiting values fulfilling a modified Gibbs' criterion. The modification stems from the fact that one has to take into account the change of the contact angle of the nanodroplet with its volume. We show that the spatial extent of the region within which the apparent contact angle changes significantly is of the order of ten fluid particle diameters and thus remains mesoscopic.  We summarize and discuss our results in Sec.\,V.


\section{Model}
In order to determine the effective interface Hamiltonian for an interface separating a liquid-like layer adsorbed on a substrate from the bulk gas phase we employ classical density functional theory (DFT). The corresponding grand canonical density functional $\Omega ([\rho(\rr)];T,\mu)$ is a function of the temperature $T$ and the chemical potential $\mu$, and it is a functional of the the spherically symmetric interparticle pair potential $\tilde w(r)$ and of the external potential $V_{ext}(\rr)$ encoding the influence of the substrate. The interparticle potential $\tilde w(r)$ is split into a short-ranged repulsive part $w_{hs}(r)$ and an attractive part $w(r)$:
\beq
	\tilde w(r) &=& w_{hs}(r)+w(r) \ .
\eeq  
Two models of the attractive part will be discussed: a short-ranged Yukawa-type potential and a long-ranged van der Waals potential.

We adopt a simple random phase approximation for the density functional \cite{Evans1979,Napiorkowski1993,Napiorkowski1994,Napiorkowski1995}:
\begin{align} 
 \begin{split}
 \Omega ([\rho(\rr)];T,\mu) =&  \int \dd^3 r  f_{hs}(\rho(\rr)) \\ & + \frac{1}{2} \int \dd^3 r \!\! \int \dd^3 r' w(|\rr-\rr'|) \rho(\rr) \rho(\rr') \\
 	& + \int \dd^3 r \left( V_{ext}(\rr)-\mu \right)\rho(\rr) \, . \label{th_functional}
\end{split}
\end{align}
The equilibrium number density profile minimizes $\Omega ([\rho(\rr)];T,\mu)$. The first term on the rhs represents the free energy in the local density 
approximation of the reference fluid interacting via the short-ranged repulsive potential $w_{hs}(r)$. The external potential $V_{ext}(\rr)$ acting on a 
fluid particle located at position $\rr$ stems from its interactions with all particles forming the substrate,
\beq \label{V_ext}
V_{ext}(\rr) &=& \int_{{\cal V}_s} \dd \rr' \, \rho_s \, w_s(|\rr-\rr'|) \ , 
\eeq 
where ${\cal V}_s$ denotes the spatial region occupied by the substrate with homogeneous number density $\rho_s$. As an approximation we take $\rho(\rr)=0$ 
in that spatial region where $V_{ext}$ is repulsive; in the remaining part of space $V_{ext}$ is determined by the attractive fluid-substrate interaction $w_s$. 

The thermodynamic state of the fluid is taken to be at the bulk liquid-gas coexistence line $\mu_0(T)$ and sufficiently below the critical point. This 
implies that the bulk correlation length is comparable with the diameter $\sigma$ of the fluid particle. Under these conditions the nonuniform number 
density profile $\rho(\rr)$ can be described within the so-called sharp-kink approximation 
\beq 
 \rho_{shk}({\bf R},z) = \rho_{l}\,\Theta(f({\bf R})-z)\, + \, \rho_{g}\,\Theta(z-f({\bf R})) \ , \label{profile}
\eeq 
where $\Theta(z)$ is the Heaviside function while $\rho_{l}$ and $\rho_{g}$ denote the bulk number densities of the coexisting liquid and gas phase, 
respectively. The local position of the liquid-gas interface is described in terms of the Monge parametrization  
$z=f(\RR=(x,y))$. 
Thus by invoking the sharp-kink approximation we disregard the actual smooth variation of the density profile due to 
thermal fluctuations and due to the long range of the interactions governing the system which give rise to so-called van 
der Waals tails \cite{Napiorkowski1989,Dietrich1991}. For a finite system the density functional $\Omega ([\rho(\rr)];T,\mu)$ in 
Eq.\,(\ref{th_functional}), evaluated at $\rho(\rr) = \rho_{shk}({\bf R},z)$, can be systematically decomposed into a sum of bulk, 
surface, line, etc. contributions \cite{Bauer2000,Getta1998,Merath2008}. 

For a system which is translationally invariant along, say, the $y$-direction, the $f$-dependent surface contribution to the density 
functional is the sum of two terms:
\begin{widetext}
\begin{align} 
 \begin{split}
  \Omega_s[f] =& L_y \Big[ \Omega_{lg}([f])+ \Omega_{int}([s],[f]) \Big] \, , \label{fdep}
 \end{split}
\end{align}
where $L_y$ is the system size in the invariant direction. The first term in the bracket corresponds to the free energy functional per 
length of a free, fluctuating liquid-gas interface:
\begin{align} \label{Omega_lg}
  \Omega_{lg}([f]) = -\frac{1}{2}(\rho_l-\rho_g)^2 \int_{-L_x}^{L_x} \dd x  \int_{-L_x}^{L_x} \dd x' 
        \int_{f(x)}^{\infty} \dd z \int_{-\infty}^{f(x')} \dd z' \int_{-\infty}^{\infty} \dd y \, w(x\!-\!x',y,z\!-\!z') \, .
\end{align} 
The second term describes the effective interaction per length of the liquid-gas interface with the surface $s(x)$ of the substrate:
\begin{align} \label{def_interaction}
 \begin{split}
  \Omega_{int}([s],[f])  =
        & (\rho_l-\rho_g) \int_{-L_x}^{L_x} \dd x  \int_{-L_x}^{L_x} \dd x'  
        \int_{f(x)}^{\infty} \dd z \int_{-\infty}^{s(x')} \dd z' \int_{-\infty}^{\infty} \dd y 
    \Big\{\rho_l w(x\!-\!x',y,z\!-\!z') 
       - \rho_s w_s(x\!-\!x',y,z\!-\!z') \Big\} \\
      \equiv & \int_{-L_x}^{L_x} \dd x \, \omega(x,f(x),[s],L_x) \, .
 \end{split}
\end{align} 
\end{widetext}
The function $\omega(x,f(x),[s],L_x)$ is the surface density of the interaction functional and it is called effective interface potential. 
The limit $L_x \rightarrow \infty$ is taken after the appropriate leading terms proportional to $L_x$ are extracted from the above expressions 
(see for example, c.f., Eq.\,(\ref{full_functional})).

For small undulations $|f'(x)| \ll 1$ the expression in the bracket in Eq.\,(\ref{fdep}) can be approximated by its local form which is called 
the local effective interface Hamiltonian of the system:
\begin{align} \label{hamiltonian}
\mathscr{H}[f] &=& \int_{-L_x}^{L_x} \dd x \, \Big\{ \gamma \sqrt{1+f'(x)^2}+\omega(x,f(x),[s])\Big\}\, .
\end{align}
Within the present approximation the surface tension coefficient of the liquid-gas interface $\gamma$ is given by
\beq \label{def_tension}
 \gamma =  -\frac{1}{2} (\rho_l - \rho_g)^2 \, \pi \int_0^\infty \dd  r \, r^3 w(r) \, .
\eeq


\section{Liquid-gas interface close to an apex-shaped substrate \label{sec_apex}}

\begin{figure}[htb] 
 \begin{center}
  \includegraphics{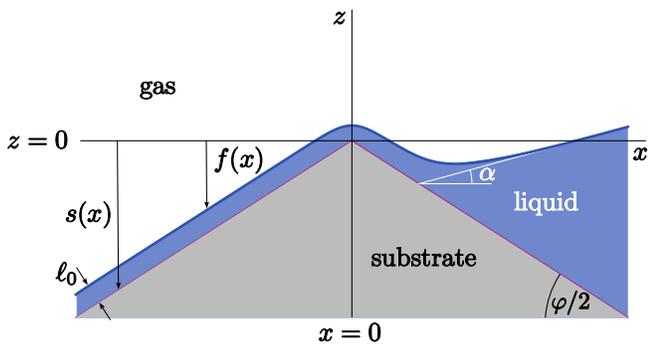}
 \caption{Schematic profile of a liquid-gas interface at an apex-shaped substrate with a characteristic angle $\varphi$. The liquid layer 
thickness $\ell_0 = \cos (\varphi/2) f(x\to -\infty)$ on the left hand side of system is finite and the interface detaches from 
the substrate with an angle $\alpha$ on the right hand side of the system. The shapes of the interface and of the substrate surface, $f(x)$ 
and $s(x)$, respectively, are measured relative to the plane $z=0$ running through the apex horizontally. The system is translationally 
invariant in $y$-direction. \label{fig_apex}}
 \end{center}
\end{figure}

In this section we aim at finding the equilibrium shape $\bar f(x)$ of the liquid-gas interface which minimizes the effective Hamiltonian 
(Eq.\,(\ref{hamiltonian})) for an apex-shaped substrate (see Fig.\,\ref{fig_apex}). 
The adsorption of a liquid phase at this kind of a substrate has been investigated in the context of wetting phenomena \cite{Parry2003}. We  
shall focus on the thermodynamic states below the wetting temperature at the bulk liquid-gas phase coexistence line. We consider configurations 
which attain a finite width $\ell_0$ at the far left hand side of the apex and detach from the substrate with an angle $\alpha$ on the right 
hand side of the apex (see Fig.\,\ref{fig_apex}). We aim at determining the range of accessible angles $\alpha$ for such configurations. As a 
first step, we recall the results for a planar substrate \cite{Indekeu1992,Dobbs1993,Getta1998,Merath2008,Indekeu2010,Indekeu2011}, for which -- 
for specific choices of the fluid-fluid and the substrate-fluid intermolecular pair potentials -- one is able to carry out the whole analysis analytically. 


\subsection{Planar substrate \label{sec_flat}} 

\begin{figure}[htb] 
 \begin{center}
  \includegraphics{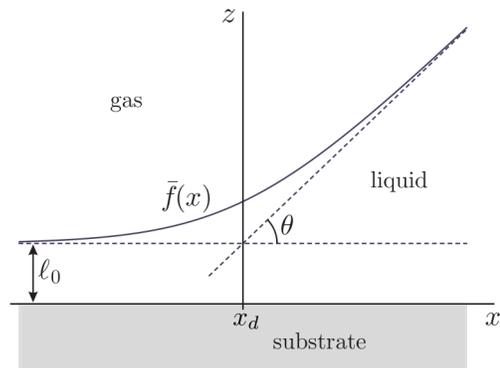}
 \caption{Schematic plot of the equilibrium liquid-gas interface $\bar f(x)$ which attains a constant value $\ell_0$ for $x \to -\infty$ 
and forms a contact angle $\theta$ with the substrate covered by the wetting film of equilibrium thickness $\ell_0$. The parameter $x_d$ 
is the lateral position of the intersection of the asymptotes \mbox{$\bar f(x\to -\infty)$} and $\bar f(x\to \infty)$. There is a family of 
solutions $\bar f (x-d)$ with the same free energy which follow from shifting the profile $\bar f(x)$ laterally by a constant $d$.  \label{schem_flat}}
 \end{center}
\end{figure}

For a \emph{p}lanar substrate the effective interface potential $\omega_p(f(x))=\omega(x,f(x),s \equiv [0])$ does not depend explicitly on $x$. Accordingly the effective interface Hamiltonian of the system (Eq.\,(\ref{hamiltonian})) is given by 
\begin{align} \label{full_functional}
\begin{split}
\mathscr{H}_p[f] = \int_{-\infty}^{\infty} \dd x \, \Big\{ & \gamma \Big(\sqrt{1+f'(x)^2}-\sqrt{1+a_p'(x)^2}\Big) \\
   & +\omega_p(f(x))-\omega_p(a_p(x))\Big\} \, ,
 \end{split}
\end{align}
where on the rhs of Eq.\,(\ref{full_functional}) the free energy per length corresponding to the asymptotic configuration (i.e., for 
$|x| \to \infty$)
\beq \label{as_full}
a_p(x) &=&  \ell_0 \, + (x-x_d) \tan \theta  \, \Theta(x-x_d)
\eeq
is subtracted so that $\mathscr{H}_p[f]$ is finite for $L_x \to \infty$. The contact angle $\theta$ (unknown a priori) is formed by 
the asymptotes in the limits $x \to -\infty$  and  $x \to \infty$ (see Fig.\,\ref{schem_flat}): 
\beq
 \theta = \lim_{x \to \infty} \arctan f'(x) \, . 
\eeq
The parameter $x_d$ determines the position of the intersection of the asymptotes, which is also unknown a priori.

The equilibrium shape $f=\bar f(x)$ of the liquid-gas interface minimizing this functional fulfills the equation
\beq \label{full_eq}
   \gamma \frac{\bar f''(x)}{{\Big[1+{\bar f'(x)}^2\Big]}^{3/2}} &=& \omega_p'(\bar f(x))  \, ,
\eeq
which after one integration leads to 
\beq \label{after_int}
 \frac{1}{\sqrt{1+{\bar f'(x)}^2}} = -\frac{\omega_p(\bar f(x))}{\gamma}+C \, ,
\eeq
where $C$ is an integration constant. Demanding $\bar f(x \to -\infty) = \ell_0$ implies  $\bar f'(x \to -\infty) = 0$, $\bar f''(x \to -\infty) = 0$, 
and   $\omega_p'(\ell_0) = 0$. According to Eq.\,(\ref{after_int}) the integration constant $C$ equals
\beq
C = 1+ \frac{\omega_p(l_0)}{\gamma} \, .
\eeq
For a profile $\bar f(x)$ diverging linearly for \mbox{$x \to \infty$} one obtains \cite{Dietrich1988}
\beq \label{cosine}
 \cos \theta = 1+ \frac{\omega_p(l_0)}{\gamma} \, .
\eeq

For a planar liquid-gas interface corresponding to a wetting film on a flat substrate, the substrate-gas surface tension coefficient 
equals the equilibrium surface free  energy density 
\beq \label{Youngslaw}
 \gamma_{sg} &=& \gamma_{sl}+\gamma+\omega_p(\ell_0) \, ,
\eeq
where $\gamma_{sl}$ denotes the substrate-liquid surface tension. Together with Eq.\,(\ref{cosine}) this renders Young's law 
\beq \label{Youngslaw2}
  \cos \theta &=& \frac{\gamma_{sg} - \gamma_{sl}}{\gamma} \, .
\eeq 


\subsubsection{Short-ranged forces \label{sec_srf}}
 
 In order to find from Eq.\,(\ref{full_eq}) the explicit expression for the equilibrium shape of the liquid-gas interface, we choose 
 \beq \label{omega_short}
   \omega_p(\ell) = 4 \gamma \Big(- a \, e^{-\ell} + e^{-2 \ell}\Big) \, 
 \eeq
as a specific model for the effective interface potential, where $\ell$ is the film thickness divided by the bulk correlation length $\xi$ in the wetting phase. We take the dimensionless amplitude $a$ within the range \mbox{$0 \leqslant a \leqslant 1$}. 
Within this model the effective interface potential attains its minimum at $\ell_0 = -\ln (a/2)$ with  $\omega_p(\ell_0) = -a^2\,\gamma$. 
The amplitude $a$ is a unique function of temperature and $a=0$ corresponds to the transition temperature of continuous wetting at 
which the equilibrium film thickness $\ell_0$ diverges and the contact angle $\theta$ vanishes ($\cos \theta = 1 -a^2$ due to Eq.\,(\ref{cosine})). 
The value $a=1$ corresponds to $\theta = \pi/2$. In Subsec.\,\ref{sec_srf} all lengths (e.g., $f$ and $x$) are measured in units of $\xi$.

Deriving the above expression (Eq.\,(\ref{omega_short})) for short-ranged intermolecular forces requires to go beyond the sharp-kink approximation (Eq.\,(\ref{profile})), which corresponds to setting the bulk correlation length equal to zero (see Refs. \onlinecite{Aukrust1985,Aukrust1987,Dietrich1988}). On the other hand, in the case of long-ranged intermolecular forces the sharp-kink approximation turns out to be not a severe one. Using this approximation in the latter case one obtains the exact expressions for the coefficients multiplying the two leading-order terms in the expansion of the corresponding effective interface potential in terms of powers of $1/\ell$ (see Ref.\,\onlinecite{Dietrich1991}). 

For weakly varying interfaces, i.e., $|f'(x)| \ll 1$, one can expand the effective Hamiltonian:
\begin{align} \label{funct_approx}
\mathscr{H}_p[f] = \int_{-\infty}^{\infty} \dd x \, \Big\{\frac{\gamma}{2}f'(x)^2+\omega_p(f(x)) - \omega_p(\ell_0) \Big\}\, .
\end{align}
The equilibrium shape of the interface $f= \bar f(x)$ minimizes the above functional and satisfies the equation
\beq \label{equi}
  \bar f''(x) &=&  \frac{\omega_p'(\bar f(x))}{\gamma} \, , 
\eeq
which upon integration yields
\beq \label{eq_sh_der}
   \frac{1}{2} \Big(f'(x)\Big)^2 &=&  \frac{\omega_p(f(x))}{\gamma} +\frac{b^2}{2} \, . 
\eeq
Here and in the following we omit the overbar indicating the equilibrium shape of the interface. The parameter $b$ is the first integration constant. 
(Note that the first term on the rhs of Eq.\,(\ref{eq_sh_der}) can be negative (compare Eq.\,(\ref{cosine})); therefore the second term must be 
positive because the lhs is positive.) A second integration renders the general solution of Eq.\,(\ref{equi}):
\begin{align} \label{solution}
f(x) = \ln \Big\{ \frac{e^{-b(x-d)}}{4b^2} \Big[ \Big(8a+e^{b(x-d)}\Big)^2-32 b^2\Big] \Big\} \, ,
\end{align}
where $d$ is the second integration constant which shifts the position of the liquid-gas interface in the horizontal direction (see Fig.\,\ref{schem_flat}). 
We put on note the property
\beq
 f(-x;-b,-d) = f(x;b,d) \, ,
\eeq
which allows one to focus on the case $b \geqslant 0$. The derivative $f'(x)$ can be rewritten in the form
\beq
 f'(x) &=& \frac{8a+e^{b(x-d)}}{2b}e^{-f(x)}-b \, .
\eeq

According to Eq.\,(\ref{solution}) for finite $d$ and nonzero $b$ the interface profile diverges for $x \to \infty$: 
\beq
 f(x\to \infty)=b(x-d)-2\ln (2b) \, . 
\eeq
On the other hand
\beq \label{minus}
 f(x \to -\infty) = - \ln \frac{a}{2} =  \ell_0 \, 
\eeq
(which implies $\ell_0 \geqslant \ln 2$), provided 
\beq \label{fixed_b}
\quad b = a \sqrt{2}  \, .
\eeq
Otherwise $ \lim_{x \to -\infty} f(x)$ is either infinite or undetermined. For $b=a \sqrt{2}$ the liquid-gas equilibrium interface profile takes the form 
\beq \label{f_b}
 f(x) = \ell_0+\ln \Big\{ 1+\frac{e^{a\sqrt{2}(x-d)}}{16a}\Big\} \, ,
\eeq
and the contact angle $\theta$ fulfills the equation
\beq \label{eq_tan}
\tan \theta = f'(x=\infty) =  a \sqrt{2} \, .
\eeq 
Equation (\ref{f_b}) implies $\cos \theta = (1+2a^2)^{-1/2} > 1 - a^2$ (compare Eq.\,(\ref{cosine})). Thus for a given $a$ the contact angle predicted by Eq.\,(\ref{funct_approx}) is smaller than the one predicted by the full model in Eq.\,(\ref{full_functional}). Accordingly, for Eq.\,(\ref{funct_approx})  
the range of angles $\theta$ accessible upon changing the parameter $0 \leqslant a \leqslant 1$ is $[0,\theta_0]$ with $\theta_0 \approx 55^\circ <90^\circ$.

\bigskip
The other solution $f(x)=-\ln (a/2)=\ell_0$ of Eq.\,(\ref{equi}) is trivial. It corresponds to $b=a \sqrt{2}$ and $d = \infty$ in Eq.\,(\ref{solution}).


\subsubsection{Boundary conditions at a finite lateral position of the three-phase contact line\label{sec_bcafp}}
One way to determine the parameters $b$ and $d$ of the equilibrium liquid-gas interface profile in Eq.\,(\ref{solution}) is to fix the value of 
the function $f(x_0)=f_0$ and of its derivative $f'(x_0)=f_0'$ at a finite position $x_0$. This implies
\beq \label{be}
 b &=& \sqrt{(f_0')^2-2 \omega_p(f_0)/\gamma} \, , 
\eeq
and
\beq \label{de}
 e^{-bd} &=& \Big[ 2 \, b \,  e^{f_0}(f_0'+b)-8 \, a \Big] e^{-b x_0} \, .
\eeq  

\begin{figure}[htb] 
 \begin{center}
   \includegraphics{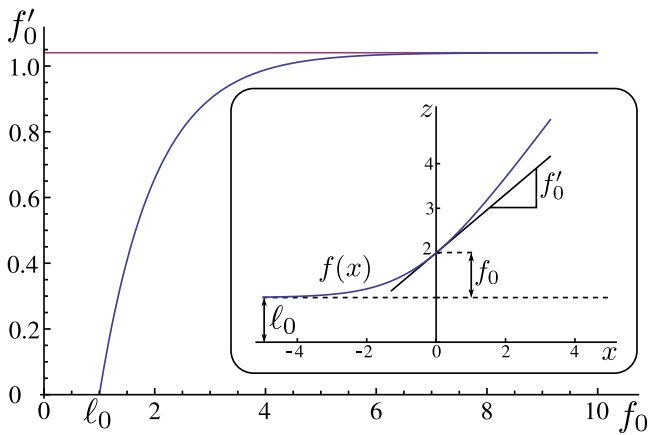}
 \caption{Dependence of the derivative $f_0'=f'(x_0)$ on the height of the interface $f_0=f(x_0)$ in cases for which the interface attains a finite value $f(x=-\infty)=\ell_0$ for $x \to - \infty$. Note that $\omega_p(f_0 \to \infty)=0$, $\omega_p(\ell_0)=- a^2 \gamma$, and $\ell_0 = -\ln (a/2)$ so that $f_0'(f_0=\ell_0)=0$. The plot corresponds to Eq.\,(\ref{omega_short}) with $a =2/e$ rendering $\ell_0=1$. The inset shows the liquid-gas configuration for the specific choice $x_0=0$ and the height of the interface $f_0=2$. Note that all lengths are measured in units of $\xi$ (see Eq.\,(\ref{omega_short})).  \label{fprime}}
 \end{center}
\end{figure}

In the previous subsection we checked that only for $b=a \sqrt{2}$ (Eq.\,(\ref{fixed_b})) the interface attains a finite value for $x \to -\infty$. 
Due to Eq.\,(\ref{be}) this leads to the relation
\beq \label{der_f0}
 f_0' &=& \sqrt{2(a^2+ \omega_p(f_0)/\gamma)} \, ,
\eeq
which is depicted in Fig.\,\ref{fprime}. 
According to Eq.\,(\ref{f_b}) the allowed values of $f_0$ are bounded from below by $\ell_0$, i.e., $f_0 \geqslant \ell_0$. 

With $b=a \sqrt{2}$ and Eqs.\,(\ref{de}) and (\ref{der_f0}) the second parameter $d$ is given by the equation
\begin{align}
  e^{-a\sqrt{2}d} = 4 a \Big[a e^{f_0} \Big(\sqrt{1\!+\! \omega_p(f_0)/(a^2 \gamma)}+1 \Big)\!-\!2 \Big] e^{- a\sqrt{2} x_0} \, .
\end{align}
For fixed $x_0$ and $f_0 \geqslant \ell_0$ the shape of the interface which attains a finite value for $x \to - \infty$ is determined uniquely. For $f_0<\ell_0$ the function $f=f(x)$ is not defined in the whole range $x \in (-\infty,\infty)$ and thus physically not acceptable.   


\subsubsection{Boundary conditions at the flat asymptote}
If one requires a finite value $f(x_0)$ for $x_0 \to -\infty$, this implies $f_0=\ell_0$ (Eq.\,(\ref{minus})), $f_0' \to 0$ (Eq.\,(\ref{der_f0})),  $b=a \sqrt{2}$ (Eq.\,(\ref{fixed_b})), and with Eq.\,(\ref{de})
\beq
   e^{-a\sqrt{2}d} &=& 4 \sqrt{2} f_0' e^{- a\sqrt{2} x_0} \label{def_d} \, .
\eeq
The value of the parameter $d$ follows from Eq.\,(\ref{def_d}), but it depends on the way in which $f_0'$ vanishes and $x_0$ approaches minus infinity.  In contrast to the case of fixing the height of the liquid-gas interface at a finite lateral position $x_0$, in the present case one obtains a whole family of solutions, which is given by Eq.\,(\ref{f_b}) and parametrized by $d$. The members of this family of solutions differ only by a constant lateral shift.


\subsection{Apex-shaped substrate \label{sec_shapex}} 

In this section we investigate the equilibrium shape of the liquid-gas interface at an apex-shaped substrate (Fig.\,\ref{fig_apex}). The substrate is translationally 
invariant in the direction perpendicular to the plane of the figure and the surface of the substrate is described by the function $s(x)=-|x| \tan (\varphi/2)$. 
We assume that the adsorbed liquid layer attains a finite width $\ell_0$ for $x \to -\infty$ and the liquid-gas interface detaches from the substrate with an angle 
$\alpha$ on the right hand side of the apex. The value of the angle $\alpha$ is not known a priori. If the detachment occurs far to the right of the apex, for the 
geometry in Fig.\,\ref{fig_apex} one expects $\alpha = \theta - \varphi/2$.
For this shape of the substrate even for the short-ranged intermolecular pair potentials the effective interface potential cannot be obtained in an analytical form 
so that the equilibrium shape of the liquid-gas interface has to be determined numerically. 

In view of this loss of analytic advantage we now consider long-ranged interactions as they are realistic for actual fluid systems. For the attractive parts of the fluid-fluid and substrate-fluid pair potentials we take \cite{Tasinkevych2006,Tasinkevych2007,Hofmann2010} 
\begin{align}\label{VdWaals}
 w(r) = -\frac{A}{(\sigma^2+r^2)^3} \, , \quad w_s(r)=-\frac{A_s}{(\sigma_s^2+r^2)^3} \, ,
\end{align}
where $A>0$ and $A_s>0$ are the amplitudes of the interactions while $\sigma$ and $\sigma_s$ are related to the molecular sizes of the fluid and substrate particles. For this model the surface tension coefficient in Eq.\,(\ref{def_tension}) takes the form
\beq \label{eq_gamma}
\gamma &=& \frac{A \pi}{8 \sigma^2}(\rho_l-\rho_g)^2 \, .
\eeq
In the case of an \emph{ap}ex shaped substrate and for the above interparticle potentials the effective interface potential reads:
\begin{widetext}
\begin{align}
 \begin{split}
 \omega_{ap}(x,f(x)) =  (\rho_l-\rho_g)  \int_{f(x)}^\infty \dd z & \int_{-\infty}^{\infty} \dd x' \int_{-\infty}^{s(x')} \dd z' \int_{-\infty}^{\infty} \dd y' 
   \Big\{\rho_l w(x\!-\!x',y',z\!-\!z') - \rho_s w_s(x\!-\!x',y',z\!-\!z') \Big\} \, .
 \end{split}
\end{align}
This leads to the disjoining pressure
\beq \label{PI}
  \Pi_{ap}(x,z) = - \frac{\partial \omega_{ap}(x,z)}{\partial z} =   \frac{\pi}{4}(\rho_l-\rho_g)  
   \Bigg[\frac{A \rho_l}{\sigma^3} \, \hat \Pi_{ap}\Big(\frac{x}{\sigma},\frac{z}{\sigma}\Big) 
 - \frac{A_s \rho_s}{\sigma_s^3} \, \hat \Pi_{ap} \Big(\frac{x}{\sigma_s},\frac{z}{\sigma_s}\Big) \Bigg] \, , 
\eeq 
where
\begin{align}
\begin{split}
\hat \Pi_{ap}(x,z) =&    \arctan \left[\left(\sqrt{x^2\!+\!z^2\!+\!1}\!+\!x\right) \tan \frac{\varphi }{4} \!+\! z\right] 
          - \arctan\left[\left(\sqrt{x^2\!+\!z^2\!+\!1}\!+\!x\right) \cot \frac{\varphi }{4}\!+\!z\right] \\
 & +\frac{ \cos \frac{\varphi}{2}}{\sqrt{1+x^2+z^2} } 
   \ \frac{2 x^2 \sin \frac{\varphi}{2} -\left[\left(x^2+z^2\right) \sin \frac{\varphi}{2} -z \sqrt{1\!+\!x^2\!+\!z^2}\right] (1+z^2 \cos^2 \frac{\varphi}{2} -x^2 \sin^2 \frac{\varphi}{2} )}
   {\left[1+(z \cos \frac{\varphi}{2} -x \sin \frac{\varphi}{2} )^2 \right] \left[1+(x \sin \frac{\varphi}{2}+z \cos \frac{\varphi}{2} )^2\right]} 
 \, .
\end{split}
\end{align} 
\end{widetext}
Here and in the following we use the fluid-fluid interaction parameter $\sigma$ as the unit of length, thus setting $\sigma=1$. Upon introducing 
dimensionless quantities 
\beq \label{dimless}
  \hat \rho = \frac{1}{2} \Big(1-\frac{\rho_g}{\rho_l}\Big) \, , \quad \hat A = \frac{A_s \rho_s}{A \rho_l} \, , 
 \quad \hat \sigma_s = \frac{\sigma_s}{\sigma}
\eeq
Eq.\,(\ref{PI}) reduces to 
\beq \label{PI2}
  \Pi_{ap}(x,y) =  \frac{\gamma}{\hat \rho}
   \Big[ \hat \Pi_{ap} \Big(x,z\Big) - \frac{\hat A}{\hat \sigma_s^3} \hat \Pi_{ap} \Big(\frac{x}{\hat \sigma_s},\frac{z}{\hat \sigma_s}\Big) \Big] \, . 
\eeq
 

\subsubsection{Equilibrium shape of the liquid-gas interface \label{sec_appprofile}}
The equilibrium profile $f=\bar f(x)$ minimizes the effective Hamiltonian 
\begin{align} \label{ham_app}
\begin{split}
\mathscr{H}_{ap}[f] = \int_{-\infty}^{\infty} \dd x \, \Big\{& \gamma \Big(\sqrt{1+f'(x)^2}-\sqrt{1+a_{ap}'(x)^2} \\
  & +\omega_{ap}(x,f(x)) - \omega_{ap}(x,a_{ap}(x))\Big\}
\end{split}
\end{align}
of the system where the contributions from the asymptote (see, c.f., Fig.\,\ref{param_d})
\begin{align} \label{eq_asymptote}
 \begin{split}
  a_{ap}(x) \!=\!&  \left[-|x| \tan \frac{\varphi}{2} + \frac{\ell_0}{\cos (\varphi/2)} \right]\Theta(x_d-x)  \\
 &\! \!+\! \!\left[\!(x\!-\!x_d) \tan \alpha \!-\! |x_d| \tan \frac{\varphi}{2} \!+\! \frac{\ell_0}{\cos (\varphi\!/\!2)}\! \right]\! \Theta(x\!-\!x_d)
\end{split}
\end{align}
have been subtracted which renders the integral in Eq.\,(\ref{ham_app}) finite. The profile $\bar f(x)$
satisfies the Euler-Lagrange equation 
\begin{align} \label{num_eq}
   \hat \rho \frac{f''(x)}{[1+ f'(x)^2]^{3/2}} \!=\! 
     \!-\! \Big[ \hat \Pi_{ap} (x,f(x)) \!-\! \frac{\hat A}{\hat \sigma_s^3} \hat \Pi_{ap} \Big(\frac{x}{\hat \sigma_s},\frac{f(x)}{\hat \sigma_s}\Big) \Big] \, .
\end{align}
Here and in the following we again omit the overbar indicating the equilibrium shape of the interface. 

\begin{figure}[htb] 
 \begin{center}
   \includegraphics{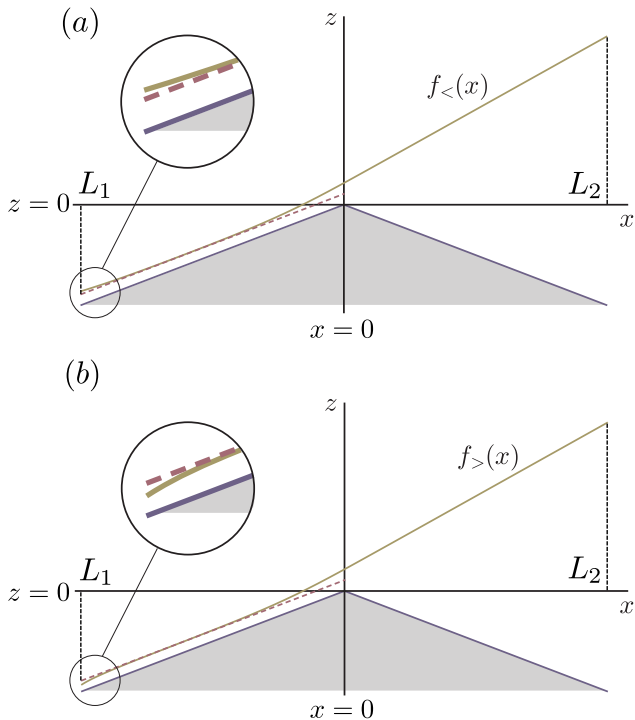}
 \caption{Schematic plots of the liquid-gas interface at an apex-shaped substrate for fixed $f_0=f(0)$ and with $f_<'<g_0'$ (a) and $f_>'>g_0'$ (b). In the former case $f_<(L_1)>a_{ap}(L_1)$, and in the latter case $f_>(L_1)<a_{ap}(L_1)$. The dashed line indicates the asymptote $a_{ap}(x)$ (Eq.\,(\ref{eq_asymptote})). These two plots correspond to case (b) in Fig.\,\ref{param_d}. \label{mniej}}
 \end{center}
\end{figure}

Equation\,(\ref{num_eq}) is integrated numerically. This is a second-order differential equation so that fixing the value of the function $f_0=f(0)$ and 
its derivative $f_0'=f'(0)$ at a certain point, say $x=0$, leads to a unique solution, similarly as discussed for the flat substrate (see Sec.\,\ref{sec_bcafp}). 
We search for solutions $f(x)$ which attain $a_{ap}(x)$ for $x \to -\infty$. There it corresponds to the same thickness $\ell_0$ of the liquid layer as for 
the corresponding wetting film on a planar substrate (see Fig.\,\ref{fig_apex} and, c.f., Fig.\,\ref{param_d}). The solutions of Eq.\,(\ref{num_eq}), which satisfy this boundary condition, 
will be called $g(x)$. In order to find them we proceed as follows:  
\begin{enumerate}
 \item fix $f_0$ and $f_0'$ at certain values $(f_0>0, \, f_0' \geqslant 0)$;
 \item integrate Eq.\,(\ref{num_eq}) numerically within the range \mbox{$x \in [L_1,L_2]$}, where $x=L_1<0$ and $x=L_2>0$ are the numerically imposed limits of the system size on the left and the right hand side, respectively;
 \item compare $f(L_1)$ with $a_{ap}(L_1)$ and $f'(L_1)$ with $a_{ap}'(L_1) = \tan (\varphi/2)$;
 \item if the differences $|f(L_1)/a_{ap}(L_1)-1|$ and $|f'(L_1)/a_{ap}'(L_1)-1|$ are not small enough we return to step 1 with a different choice of $f_0'$, but the same choice of $f_0$.
\end{enumerate}
Typical values of $|L_1|$ are in the range of tens of $\sigma$ whereas $L_2$ can be very large, e.g., $10^6 \sigma$. The necessity 
to consider only relatively small values of $|L_1|$ (as compared with $L_2$) is related to the fact that a significantly higher accuracy 
is needed to solve the differential equation in the region where the liquid-gas interface is close to the substrate. Due to limited numerical 
accuracy and due to finite system sizes we are not able to find (for a fixed value $f_0$) the value of the derivative $f_0'$ which renders 
exactly the function $g(x)$ with $g_0=f(0)$ and $g_0'=f'(0)$. What can be achieved numerically is to find values 
$f_0'=f_<'<g_0'$ and $f_0'=f_>'>g_0'$ rendering solutions which for sufficiently large and negative $x$ follow the asymptote $a_{ap}(x)$ 
and differ only slightly from it in the vicinity of $x=L_1$, as depicted in Fig.\,\ref{mniej}. These functions are called $f_<(x)$ and $f_>(x)$, respectively. 
The values of $f(L_1)$ and $f(L_2)$ change continuously with $f_0'$ so that the contact angle $\alpha = \arctan g'(L_2)$ is bounded from below by 
$\alpha_< = \arctan f_<'(L_2)$ and from above by $\alpha_> = \arctan f_>'(L_2)$, i.e., $\alpha_< \leqslant \alpha \leqslant \alpha_>$.  

\subsubsection{Gibbs' criterion}
\begin{figure}[htb] 
 \begin{center}
   \includegraphics{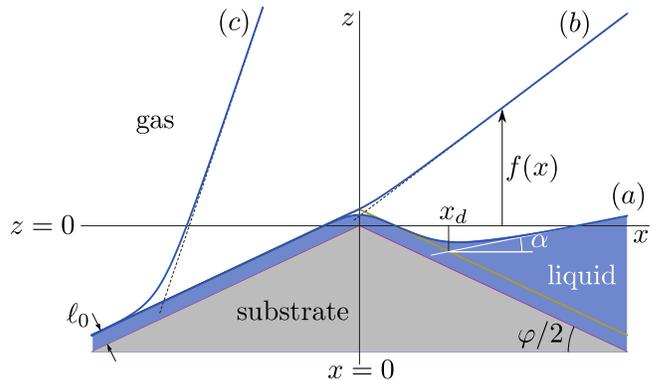}
 \caption{Schematic shapes of liquid-gas interfaces at an apex-shaped substrate. The parameters $\alpha$ and $x_d$ characterize the equilibrium liquid-gas 
interface $f(x)$ and are defined in the main text. Various choices for $f_0=f(0)$ lead to the cases $(a)$, $(b)$, and $(c)$. For $x_d \to \infty$ the local 
contact angle $\alpha+\varphi/2$ on the far right side approaches the contact angle $\theta$ on a planar substrate. \label{param_d}}
 \end{center}
\end{figure}

In order to access Gibbs' criterion we analyze the dependence of the results of the procedure described in the previous subsection on the choice of the value of $f_0$. As in the case of a planar substrate (Sec.\,\ref{sec_flat}) there exists a minimal value of $f_0$, denoted as $f_{0}^{min}$, such that solutions $g(x)$ of Eq.\,(\ref{num_eq}) exist for $f_0 \geqslant f_0^{min}$. The solution $g^{min}(x)$ corresponding to $f_{0}^{min}$ is symmetric: $g^{min}(-x) = g^{min}(x)$. For solutions $g(x)$ corresponding to $f_0 > f_{0}^{min}$ we define the contact angle $\alpha = \arctan g'(L_2)$ and the parameter $x_d$, where $x_d$ fulfills the equation $s(x_d)+\ell_0/\cos (\varphi/2) =g'(L_2)(x_d-L_2)+g(L_2)$. The parameter $x_d$ characterizes the position at which the liquid-gas interface detaches from the substrate. This is defined as the intersection of the corresponding asymptotes (see Fig.\,\ref{param_d}). 

\begin{figure}[htb] 
 \begin{center}
   \includegraphics{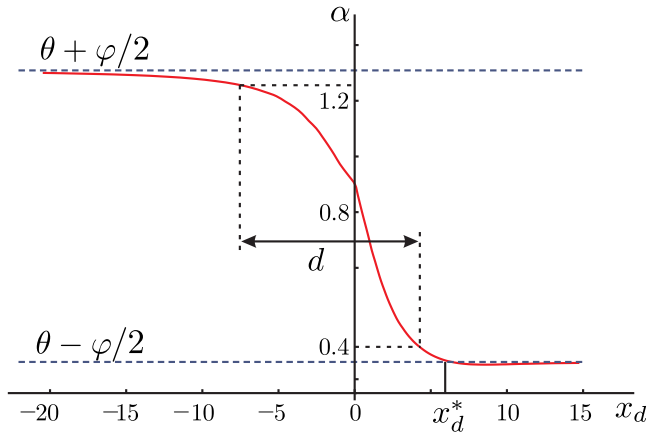} 
 \caption{Dependence of the contact angle $\alpha$ on the parameter $x_d$ characterizing the position of the liquid-gas interface detachment from the substrate (see Fig.\,\ref{param_d}). The horizontal dashed lines from bottom to top indicate the angles $\theta-\varphi/2$, $\theta-0.9\,\varphi/2$, $\theta+0.9\,\varphi/2$, and $\theta+\varphi/2$, respectively, where $\theta$ is the contact angle on a planar substrate and $\pi-\varphi$ is the opening angle of the apex. The quantity $d$ measures the width of the region within which the contact angle changes between the values  $\alpha = \theta-0.9\,\varphi/2$ and $\alpha = \theta+0.9\,\varphi/2$ and $x_d^*$ is the position above which $\alpha<\theta-\varphi/2$. These data correspond to $L_1=-30 \sigma$, $L_2=10^6 \sigma$, $\varphi=\pi/3$, $\hat \sigma=0.5$, $\hat A = 0.82$, and $\hat \rho = 0.023$ such that the effective interface potential renders $\ell_0=2 \sigma$ and $\theta=\pi/4$. All lengths are measured in units of $\sigma$. The break in slope of $\alpha(x_{d})$ at $x_{d}=0$ is caused by the discontinuity in the derivative of the function $a_{ap}(x)$ (Eq.(\ref{eq_asymptote})) which enters the definition of the point $x_{d}$. 
 \label{crit}}
 \end{center}
\end{figure}

Upon increasing $f_0 > f_{0}^{min}$ the contact angle $\alpha(f_0)$ increases and the parameter $x_d(f_0)$ decreases, i.e., the three-phase contact line approaches 
the apex. Changing the value of $f_0$ enables one to plot the dependence of the contact angle $\alpha$ on the parameter $x_d$ (Fig.\,\ref{crit}). 
The difference between $\alpha_>$  and $\alpha_<$ is so small that the error bars of $\alpha$ are not visible on the present scale.  

In the case of a planar substrate the free energies corresponding to the asymptotic configurations (Eq.\,(\ref{as_full})) are the same for each 
interface profile. Therefore the task of finding the equilibrium configuration, i.e., the profile with the lowest free energy (Eq.\,(\ref{full_functional})), 
is posed well. In the case of the apex-shaped substrate, for liquid-gas configurations fulfilling Eq.\,(\ref{num_eq}) with appropriate boundary conditions, 
the free energies of the corresponding asymptotic configurations, which represent different constraints,  are different (Eqs.\,(\ref{ham_app}) 
and (\ref{eq_asymptote})). Thus, comparing free energies corresponding to different configurations amounts to compare free energies 
characterizing different constraints. These free energies as function of $x_d$ can be interpreted as the potential of the effective 
interaction between the three-phase contact line and the apex.

For \mbox{$x_d \to \pm \infty$} the local contact angle \mbox{$\alpha=\arctan g'(L_2)$} tends to its limiting values $\theta \mp \varphi/2$ from below, which are those expected from Gibbs' criterion for this geometry (compare Eq.\,(\ref{gibbs_criterion}) which holds for the geometry shown in Fig.\,\ref{fig_gibbs}). For $x_d>x_d^*$ the local contact angles are, slightly, smaller then $\theta-\varphi/2$ (Fig.\,\ref{crit}). The spatial extent of the region within which the contact angle $\alpha$ changes significantly can be chosen as the region of $x_d$ where 
$\theta-0.9\,\varphi/2 \leqslant \alpha \leqslant \theta+0.9\,\varphi/2$. For $\varphi=\pi/3$ and for parameters of the effective interface potential rendering $\ell_0=2 \sigma$ and $\theta=\pi/4$, this width equals $d\approx 11\, \sigma$ and thus it is mesoscopic.

\section{Sessile droplets on trapezoidal substrates \label{sec_droplet}}

\subsection{Planar substrate}
\begin{figure}[htb]
 \begin{center}
  \includegraphics{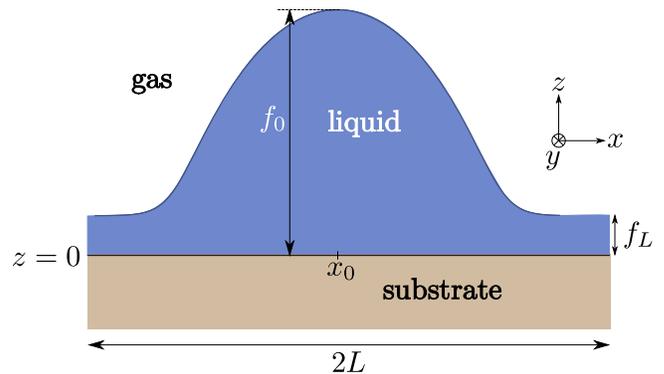}  
  \caption{Schematic equilibrium shape of a ridgelike liquid nanodroplet deposited on a planar substrate. The system is translationally invariant in 
$y$-direction and has a finite lateral extent $|x| \leqslant L_x \equiv L $. \label{fig_schemdrop}}
 \end{center}
\end{figure}

As preparatory work, in this subsection we discuss the shapes of interfaces characterizing sessile droplets on planar substrates 
(see Fig.\,\ref{fig_schemdrop}).
We assume that the system under consideration has a finite extent in $x$-direction, $|x| \leqslant L_x \equiv L$, and is translationally invariant in the 
$y$-direction. The shape of the interface is described by a function $f=f(x)$. The total volume $V_{tot}/L_y$ of liquid in the system per length $L_y$ is fixed. 
$L_{y}$ denotes the size of the system in $y$ direction.  

Within our mesoscopic description for weakly varying liquid-gas interfaces the corresponding effective Hamiltonian is given by (compare Eq.\,(\ref{funct_approx}))
\beq \label{ham_drop}
\mathscr{H}_p[f] &=& \int_{-L}^{L} \dd x \, \Big\{\frac{\gamma}{2}f'(x)^2+\omega_p(f(x))\Big\}\, ,
\eeq
where the limits of the $x$-integration reflect the finite extent of the system. The equilibrium shape $\bar f(x)$ of the interface  minimizes the functional
\beq
\mathscr{H}^*_p[f] &=& \mathscr{H}_p[f] - \lambda V_{tot}/L_y \, ,
\eeq
where $\lambda$ is a Lagrange multiplier, and 
\beq 
V_{tot} &=& L_y \int_{-L}^{L} \dd x \, f(x) \, .
\eeq
The equilibrium profile satisfies the differential equation 
\beq \label{sol_sk}
  \gamma \bar  f''(x) &=&  \omega_p'(\bar f(x)) - \lambda   \, .
\eeq
In the following we again omit the overbar denoting the equilibrium configuration. In $x$-direction we impose Neumann and periodic boundary conditions: 
\beq
 f'(-L) = f'(L) = 0 \, , \quad  f(-L) = f(L) = f_L  \, ,
\eeq
with the thickness $f_L$ not fixed a priori. The conditions $f'(-L)=f'(L)=0$ can be realized by vertical sidewalls at $x = \pm L$ exhibiting a contact angle of $90^{\circ}$.

\begin{figure}[htb] 
 \begin{center}
 \includegraphics{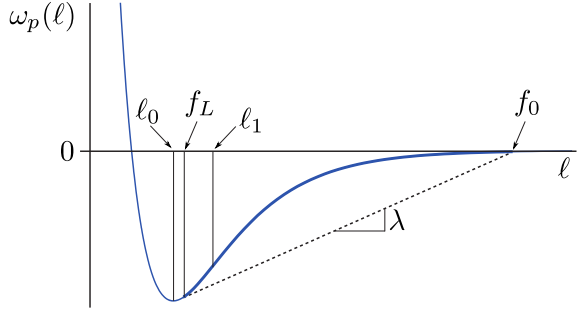} 
 \caption{Schematic plot of the effective interface potential considered here in the context of discussing the shape of small droplets 
(Eqs.\,(\ref{sol_fL_sk}) and (\ref{L})). The slope of the dotted line connecting the points $(f_0,\omega_p(f_0))$ and $(f_L,\omega_p(f_L))$  
equals the Lagrange multiplier $\lambda$ (see Eq.\,(\ref{f0_sk})). \label{fig_constr_sk}}
 \end{center}
\end{figure}

Integrating Eq.\,(\ref{sol_sk}) renders 
\beq
   \frac{\gamma}{2} f'(x)^2 &=&  \omega_p(f(x)) - \lambda \, f(x) - C \, ,
\eeq
where the integration constant $C$ is determined by the boundary conditions:
\beq
 C = \omega_p(f_L) - \lambda \, f_L \, ,
\eeq
which leads to 
\beq \label{sol_fL_sk}
   \frac{\gamma}{2} f'(x)^2 &=&  \omega_p(f(x)) - \omega_p(f_L) - \lambda ( f(x) -f_L) \, .
\eeq
We examine effective interface potentials  $\omega_p(\ell)$ with a minimum at $\ell=\ell_0$, $\omega(\ell \to \infty) = 0^-$, and one 
inflection point at $\ell=\ell_1 > \ell_0$ (see Fig.\,\ref{fig_constr_sk}). 

We search for solutions \mbox{$f(x) \geqslant f_L$}, such that \mbox{$f''(x=\pm L)>0$} and that there is one $x_0 \in (-L,L)$ for which $f'(x_0)=0$; 
we denote the maximum value of the function $f(x)$ as $f_0 \equiv f(x_0)$. As a result one obtains from Eq.\,(\ref{sol_sk}) the relation 
$\lambda \leqslant \omega_p'(f_L)$ and from Eq.\,(\ref{sol_fL_sk}) one has 
\beq \label{f0_sk}
  \lambda = \frac{\omega_p(f_0)-\omega_p(f_L)}{f_0-f_L}  \, .
\eeq
From the structure of $\omega_p(f)$ one infers $\lambda>0$. Thus the slope of the line connecting the points $(f_L,\omega_p(f_L))$ and $(f_0,\omega_p(f_0))$ 
is equal to the Lagrange multiplier $\lambda$. In order to fulfill the condition $\lambda \leqslant \omega_p'(f_L)$ this line segment must be located 
below the graph of the effective interface potential (see Fig.\,\ref{fig_constr_sk}). This implies two restrictions: one for the thickness of the liquid layer 
at the boundaries $f_{L}>\ell_{0}$, and the other for the maximum value of the function $f_{0} > \ell_{1}$.

Integrating Eq.\,(\ref{sol_fL_sk}) one obtains
\begin{align} \label{L}
 L = \sqrt{\frac{\gamma}{2}} \int_{f_L}^{f_0} \dd z  \Big[ \omega_p(z) - \omega_p(f_L) - \lambda (z -f_L)\Big]^{-1/2} \, . 
\end{align}
On the other hand, by rearranging the limits of integration on the rhs of Eq.\,(\ref{L}) it can be expressed as $L-x_0$ which renders $x_0=0$. In addition, the function $f(-x)$ also fulfills Eq.\,(\ref{sol_fL_sk}) so that in the following we consider functions $f(x)$ which are symmetric with respect to $x=0$. The excess volume  
\beq
  V_{ex} \equiv V_{tot} - 2 L \, L_y \, f_L
\eeq
of the adsorbed liquid is given by (see Eq.\,(\ref{sol_fL_sk}))
\begin{align} \label{volume_sk}
\begin{split}  
V_{ex} =& 2 L_y \int_{-L}^0 \dd x \, \Big[f(x)-f_L \Big]  \\
   =& \sqrt{2} L_y \sqrt{\gamma} \,  \int_{f_L}^{f_0} \dd z \, \frac{z-f_L}{ \sqrt{ \omega_p(z) - \omega_p(f_L) - \lambda (z -f_L)}} \, .
\end{split}
\end{align}

By combining Eqs.\,(\ref{f0_sk}) -- (\ref{volume_sk}) with a given volume $V_{tot}$ and a size $L$ one determines  
the quantities $f_L$, $f_0$, $\lambda$; integrating Eq.\,(\ref{sol_fL_sk}) gives the shape of the equilibrium liquid-gas interface in the form (for $x>0$)
\begin{align} \label{inverted_x}
 x (f) = L - \sqrt{\frac{\gamma}{2}} \int_{f_L}^{f} \dd z \frac{1}{\sqrt{ \omega_p(z) - \omega_p(f_L) - \lambda (z -f_L)}} \, .
\end{align}

In the following, for the surface tension (Eq.\,(\ref{eq_gamma})) and for the effective interface potential we adopt the expressions following from the 
sharp-kink approximation for the density functional using the fluid-fluid and substrate-fluid pair potentials given by Eq.(\ref{VdWaals}):
\beq
\omega_p(\ell) &=&  \frac{\gamma}{\hat \rho} \Big[ \hat \omega_p \Big(\frac{\ell}{\sigma}\Big) 
      \, -\, \frac{\hat A}{\hat \sigma_s^2} \, \hat \omega_p\Big( \frac{\ell}{\sigma} \frac{1}{\hat \sigma_s}\Big)\Big] \, , \label{omega_long}
\eeq
where $\hat \omega_p(\ell)=\ell \arctan(1/\ell)-1$, and the three parameters $\hat \rho$, $\hat \sigma_s$, $\hat A$ are given by Eq.\,(\ref{dimless}). Instead of using the parameters $\hat A$ and $\hat \rho$, we use the width $\ell_0$ minimizing the effective interface potential $\omega_p(\ell)$ and the Young contact angle given by $\cos \theta = 1+\omega_p(\ell_0)/\gamma$. For a given $\hat \sigma_s$ the relation between $(\hat A,\hat \rho)$ and $(\ell_0,\theta)$ is unique.


\subsubsection{Stability of the droplet}
In addition to the droplet-like solution, Eq.\,(\ref{sol_fL_sk}) has the trivial, flat solution $f(x) \equiv f_L$ with zero excess volume and nonzero total volume. 
We fix the lateral size $L$ and check which of these solutions has the lower free energy (Eq.\,(\ref{ham_drop})), and thus corresponds to the stable 
interface configuration.

\begin{figure}[htb]
 \begin{center}
  \includegraphics{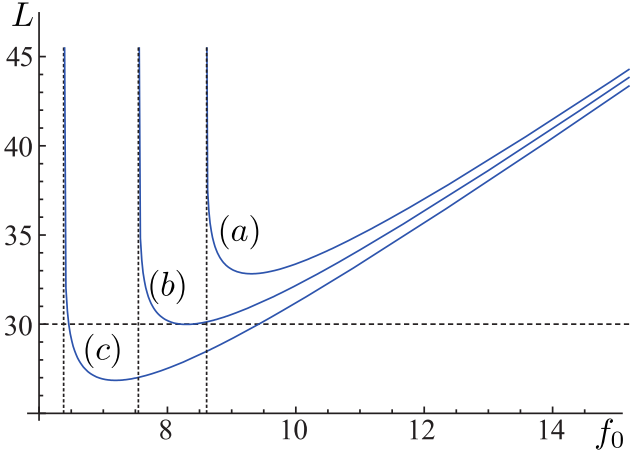} 
 \caption{Dependence of $L$ on $f_0$, according to Eqs.\,(\ref{L}) and (\ref{f0_sk}), for three different choices of 
$f_L$: (a) $f_L = 2.1$, \mbox{(b) $f_L=f_L^*=2.119$}, (c) $f_L=2.15$. For each choice of $f_L$, $f_0$ is larger than a minimal value $f_0^*(f_L)$ at which $L$ diverges 
as $L(f_0 \to f_0^*) \sim - \log (f_0-f_0^*)$. 
The surface tension coefficient and the potential parameters are chosen such that $\hat \sigma_{s}=0.5$, $\ell_0 = 2$, and $\theta=\pi/4$. 
If one prescribes the width $2L$ of the system (e.g., $2L=60$, dashed horizontal line), depending on the choice of $f_L$ there are two $(c)$, one $(b)$, 
or no corresponding values for $f_0$. For $L=30$ the threshold value is $f_L^*=2.119$ $(b)$. All lengths are given in units of $\sigma$. \label{fig_Lf0fL}}
 \end{center}
\end{figure}

For a finite system with a prescribed fixed $L$ the droplet solution does not exist for arbitrary $f_L>\ell_0$. This is caused by the 
boundary conditions $f'(-L)=f'(L)=0$. If $f_L$ is sufficiently small, i.e., $f_L<f_L^*(L)$ (where the threshold value $f_L^*$ depends also on the 
parameters of the effective interface potential), there is no $f_0$ which, according to Eqs.\,(\ref{L}) and (\ref{f0_sk}), would correspond to the 
prescribed fixed $L$ (Fig.\,\ref{fig_Lf0fL}). 
For $f_L>f_L^*(L)$ there are two values of $f_0$ which correspond to the same fixed $L$ (Fig.\,\ref{fig_Lf0fL}). As expected intuitively, one can show 
that the configuration corresponding to the smaller value of $f_0$ has always the lower free energy. 
   
\begin{figure}[htb]
 \begin{center}
     \includegraphics{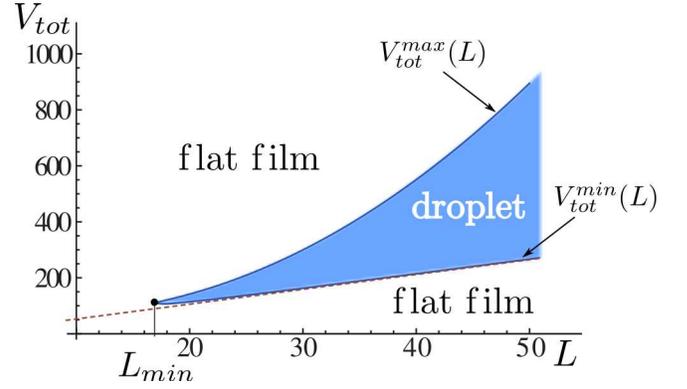} 
  \caption{Morphology phase diagram in terms of the variables $(L,V_{tot})$ displaying the coexistence lines between the phases with a droplet and 
a flat film configuration, respectively. The upper ($V_{tot}=V_{tot}^{max}(L)$) and the lower ($V_{tot}=V_{tot}^{min}(L)$) solid lines correspond to 
the droplet configurations with $f_L=f^*_L$ and $f_L=\ell_1$, respectively. The dashed straight line shows the volume of the flat film configuration 
with the thickness \mbox{$f_L \equiv \ell_1$}. $L_{min}$ denotes the smallest lateral width of the system for which the droplet configuration can exist. The 
surface tension coefficient and the potential parameters are chosen such, that $\hat \sigma_{s}=0.5$, $\ell_0 = 2$, and $\theta=\pi/4$. Note that all 
lengths are measured in units of $\sigma$ and $V_{tot}$ is measured in units of $L_y \sigma^2$. \label{fig_VminVmax}}
 \end{center}
\end{figure}

On the other hand, for a given value of $L$, the quantity $f_L$ is bounded also from above: $f_L^* \leqslant f_L \leqslant \ell_1$, rendering upper 
($V_{tot}^{max}(L)$) and  lower ($V_{tot}^{min}(L)$) bounds for the total volume for which the equilibrium droplet configuration exists (Fig.\,\ref{fig_VminVmax}).
One can check that for $f_L>\ell_1$ the droplet configuration has a larger free energy than the flat film configuration with the same total volume.
The corresponding phase diagram is presented in Fig.\,\ref{fig_VminVmax}.
   
It is worth noticing that there exists a smallest value $L_{min}$ below which the droplet configuration cannot exist. Upon increasing $V_{tot}$ 
for $L>L_{min}$ the droplet configuration forms discontinuously at $V_{tot}= V_{tot}^{min}(L)$ and ceases to exist for $V_{tot}= V_{tot}^{max}(L)$, 
also in a discontinuous way. The transition values $V_{tot}^{max}$ and $V_{tot}^{min}$ correspond to $f_L$ being equal to $f_L^*$ and to $\ell_1$, 
respectively. For $L<L_{min}$ the system extension $2 L$ turns out to be too small to accommodate the droplet and to simultaneously fulfill the 
boundary conditions $f'(-L)=f'(L)=0$. For $L \to \infty$ the minimal volume increases linearly as function $V_{tot}^{min}(L) = 2 \, L \, \ell_1$ 
(so that for large $L$ in Fig.\,\ref{fig_VminVmax} the lower bound of the droplet phase approaches the red dashed line) and the first-order character of the transition between the flat film and the droplet configuration weakens and becomes continuous at $L=\infty$.


\subsubsection{Macroscopic system} 
In this section we recall the analyses of a liquid droplet adsorbed at a flat, unbounded substrate which can spread over the whole macroscopic extension of the substrate \mbox{($L \to \infty$)} \cite{Yeh1999a,Yeh1999b,Bertozzi2001,Glasner2003,Gomba2009,Weijs2011}. If one places a droplet onto an infinitely extended, flat liquid film, after some time the droplet will disappear into the film without changing the thickness of the latter, because the droplet volume is vanishingly small compared with the total liquid volume. Accordingly, in theory here we fix the excess volume of the droplet above the flat film. 
In practice, providing an experimental setup which on one hand mimics infinite substrate extensions and on the other hand allows for the persistence of a droplet configuration seems to be rather challenging.

The effective Hamiltonian for such a liquid-gas interface $f=f(x)$ with a \emph{d}rop is given by
\begin{align}
 \begin{split}
  \mathscr{H}_{pd}[f] = \int_{-\infty}^{\infty} \dd x \, \Big\{&\gamma \Big( \sqrt{1+ f'(x)^2}-1 \Big) \\
  & +\omega_p(f(x))-\omega_p(\ell_\infty)\Big\} \, , 
\end{split}
\end{align}
where $\ell_\infty$ is the, a priori unknown, height of the equilibrium liquid-gas interface at infinity. The equilibrium profile $f= \bar f(x)$ minimizes the functional
\beq
\mathscr{H}^*_{pd}[f] &=& \mathscr{H}_{pd}[f] - \lambda V_{ex}/L_y \, ,
\eeq
where $\lambda$ is a Lagrange multiplier, and the excess volume is given by
\beq 
V_{ex} &=& L_y \int_{-\infty}^{\infty} \dd x \, \Big[ f(x) - \ell_\infty \Big] \, .
\eeq
For $x \to \pm \infty$ one has $f'(x)=f''(x)=0$ so that there the interface profile attains a certain finite height $\ell_\infty \geqslant \ell_0$. We recall that $\ell_0$ minimizes the effective interface potential, i.e., $\omega_p'(\ell_0)=0$, and it equals the equilibrium thickness of a planar liquid film adsorbed at a flat substrate in a grand canonical system in contact with a reservoir.

 The equilibrium profile $f(x)$ (here and in the following we omit the overbar indicating the equilibrium profile) satisfies the equation (compare Eq.\,(\ref{sol_sk})) 
\beq \label{sol}
  \gamma \frac{f''(x)}{(1+f'(x)^2)^{3/2}} &=&  \omega_p'(f(x)) - \lambda   \, , 
\eeq
which gives
\beq
   \gamma\frac{1}{\sqrt{1+ f'(x)^2}} &=&  -\omega_p(f(x)) + \lambda \, f(x) + C \, . 
\eeq
The integration constant $C$ and the Lagrange multiplier $\lambda$ follow from the boundary conditions at infinity:
\beq \label{eq_lambda}
 \lambda &=& \omega_p'(\ell_\infty) 
\eeq
and
\beq
 C &=& \gamma + \omega_p(\ell_\infty) - \omega_p'(\ell_\infty) \, \ell_\infty \, .
\eeq
The equation for the liquid-gas interface configuration reads:
\begin{align} \label{solf_fL}
 \begin{split}
   \gamma\frac{1}{\sqrt{1+ f'(x)^2}}  \,=\,&  \gamma- \omega_p(f(x)) + \omega_p(\ell_\infty)  \\ & + \omega_p'(\ell_\infty) ( f(x) -\ell_\infty) \, .
\end{split}
\end{align}
 
\begin{figure}[htb] 
 \begin{center}
  \includegraphics{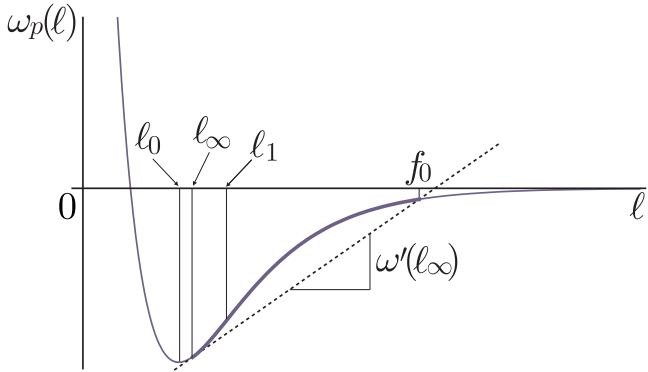}  
 \caption{Schematic plot of the effective interface potential. The values of the function $f(x)$ describing the equilibrium interface configuration lie in the range 
  \mbox{$f(x) \in [\ell_\infty,f_0]$}; $\ell_\infty$ and $f_0$  are the solutions of the equation 
  \mbox{$\omega_p(\ell) = \omega_p'(\ell_\infty)(\ell-\ell_\infty)+\omega_p(\ell_\infty)$} (dotted line). The widths $\ell_0$ and $\ell_1$ denote the abscissa of the minimum and of the inflection point of the effective interface potential, respectively. \label{fig_constr}}
 \end{center}
\end{figure}

We search for solutions $f(x) \geqslant \ell_\infty$. The values of the function $f(x)$ describing the equilibrium interface configuration lie in the range $[\ell_\infty,f_0]$ (Fig.\,\ref{fig_constr}). The boundaries $\ell_\infty$ and $f_0$ of this interval are the solutions of equation $\omega_p(\ell) = \omega_p'(\ell_\infty)(\ell-\ell_\infty)+\omega_p(\ell_\infty)$. The value $f_0 = f(x_0)$ is the maximum value of the function $f(x)$. It is a function of $\ell_\infty$ given by (Eq.\,(\ref{solf_fL}))
\beq \label{f0}
  \omega_p'(\ell_\infty)  = \frac{\omega_p(f_0) - \omega_p(\ell_\infty)}{f_0 - \ell_\infty} \, .
\eeq
This means that the line tangent to the curve \mbox{$\omega_p=\omega_p(\ell)$} at the point \mbox{$\ell=\ell_\infty$} intersects the curve \mbox{$\omega_p=\omega_p(\ell)$} again at the point \mbox{$\ell=f_0$} (Fig.\,\ref{fig_constr}) \cite{Glasner2003}. The function $f(x)$ is symmetric with respect to the \mbox{$x=x_0$}, i.e., \mbox{$f(x_0+x)=f(x_0-x)$}. 

For an effective interface potential $\omega_p(\ell)$, as shown in Fig.\,\ref{fig_constr}, with $\omega_p(\ell \to 0)>0$, $ \omega_p(\ell\to \infty)= 0^-$, and one inflection point, $\omega_p''(\ell_1)=0$, the thickness $\ell_\infty$ of the liquid layer at infinity is restricted to  $\ell_0 < \ell_\infty \leqslant \ell_1$ and thus $f_0 \geqslant \ell_1$. (These inequalities follow from the fact that the dotted line in Fig.\,\ref{fig_constr} is tangent to $\omega_p(\ell)$ at $\ell=\ell_\infty$.) For $\ell_\infty=\ell_1$ one has the flat solution $f(x) \equiv \ell_\infty$. 

For weakly varying liquid-gas interfaces ($|f'(x)| \ll 1$) Eq.\,(\ref{solf_fL}) reduces to (compare Eq.\,(\ref{sol_fL_sk}) in conjunction with Eq.\,(\ref{eq_lambda})) 
\begin{align} \label{sol_fL}
 \begin{split}
   \frac{\gamma}{2}  f'(x)^2 \,=\,&  \omega_p(f(x)) - \omega_p(\ell_\infty) \\ & - \omega_p'(\ell_\infty) ( f(x) -\ell_\infty) \, .
\end{split}
\end{align}
The excess volume of the adsorbed liquid can be expressed as (compare Eq.\,(\ref{volume_sk}) in conjunction with Eq.\,(\ref{eq_lambda})) 
\begin{align} \label{volume}
  \begin{split}
  V_{ex}= & L_y  \int_{-\infty}^\infty \dd x \, \Big[ f(x) - \ell_\infty \Big] \\
    = & L_y \sqrt{2\gamma} \int_{\ell_\infty}^{f_0} \dd \! z 
          \frac{z-\ell_\infty}{\sqrt{ \omega_p(z) \!-\! \omega_p(\ell_\infty) \!-\! \omega_p'(\ell_\infty) (z \!-\!\ell_\infty)}} \, .
 \end{split}
\end{align}

Combining Eqs.\,(\ref{f0}) and (\ref{volume}) with a given expression $\omega_p(\ell)$ for the effective interface potential and a given excess volume $V_{ex}$ one is able to determine the quantities $\ell_\infty$, $f_0$; integrating Eq.\,(\ref{sol_fL}) gives the shape of the equilibrium liquid-gas interface configuration as (compare Eq.\,(\ref{inverted_x}) in conjunction with Eq.\,(\ref{eq_lambda}))
\begin{align}
 \begin{split}
 x (f)  \!=\! x_0\!+\! \sqrt{\frac{\gamma}{2}} \! \int_{f}^{f_0}\!\! \dd z  
     \frac{1}{\sqrt{\omega_p(z) \!-\! \omega_p(\ell_\infty) \!-\! \omega_p'(\ell_\infty) (z \!-\!\ell_\infty)}} \, ,
\end{split}
\end{align}
for $x > x_0$. Due to the translational invariance of the substrate the position $x_0$ is finite and arbitrary. The solution of the second order differential equation for the equilibrium shape of the interface (Eq.\,(\ref{sol})) contains two integration constants: $x_0$ and $\ell_\infty$. The latter one can be determined from the fixed excess  volume constraint (Eq.\,(\ref{volume})) once $f_0$ is expressed in terms of $\ell_\infty$ by using Eq.\,(\ref{f0}).

\begin{figure}[htb]
 \begin{center}
  \includegraphics{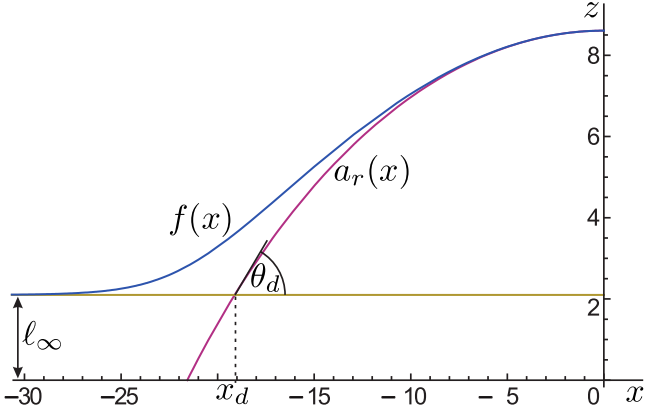}  
  \caption{Shape $f(x)$ of the ridgelike equilibrium liquid nanodroplet and the fitted arc of a circle $a_r(x)$. The contact angle for the 
droplet is denoted by $\theta_d$. The effective interface potential parameters are $\hat \sigma_s=0.5$, $\theta=\pi/4$, and $\ell_0=2$; 
$x_0$ is chosen to be $0$. The excess volume is such that $\ell_\infty = 2.1$ and $f_0 = 8.61$. All lengths are measured in units of $\sigma$. \label{fig_drop}}
 \end{center}
\end{figure}

\begin{figure}[htb] 
 \begin{center}
   \includegraphics{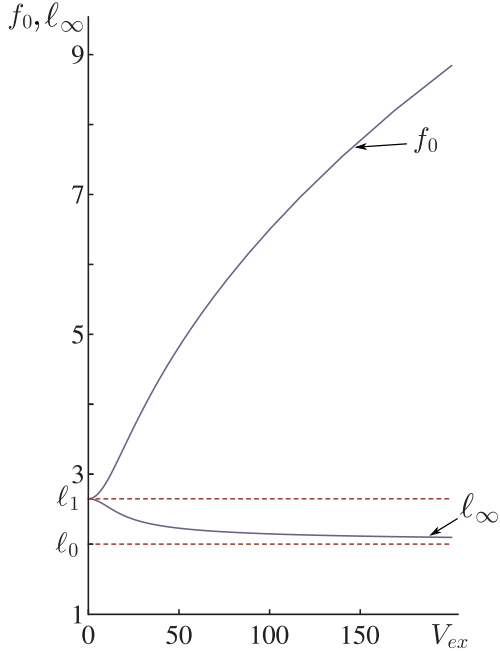} 
  \caption{Dependence of the thickness $\ell_\infty$ (lower curve) and of the maximal height $f_0$ (upper curve) on the excess volume 
$V_{ex}$ for the parameters $\hat \sigma_s=0.5$, $\theta=\pi/4$, and $\ell_0=2$. The values of $\ell_0$ and $\ell_1=2.65$ 
(compare Fig.\,\ref{fig_constr}) are marked by dashed horizontal lines. All lengths are measured in units of $\sigma$ and 
$V_{ex}$ in units of $L_y \sigma^2$. \label{fig_f0fL}}
 \end{center}
\end{figure}

The typical shape of the equilibrium liquid-gas interface is shown in Fig.\,\ref{fig_drop}. The function $f(x)$ is symmetric around the position $x_0$ of the maximum. The excess volume $V_{ex}$ of the drop uniquely determines $\ell_\infty$ and $f_0$ (see Eqs. (\ref{volume}) and (\ref{f0})). 
For $V_{ex} \to 0$ they both reach, with vanishing slope, the position $\ell_1$ of the inflection point of the effective interface potential (see Fig.\,(\ref{fig_f0fL})). For $V_{ex} \to \infty$ the maximal height $f_0$ grows without limit and $\ell_\infty$ approaches $\ell_0$. 

\subsubsection{Contact angles}                                                                                                                                                                                                                                                                                                                             
For nanodroplets the definition of the contact angle requires more care than for macroscopic drops. We define the contact angle of the 
nanodroplet as follows (Fig.\,\ref{fig_drop}): 
\begin{enumerate}
 \item  find the arc of a circle $a_r(x)=\sqrt{R^2-x^2}-R+f_0$ with the same curvature $1/R$ as the curvature of the liquid-gas interface at 
its maximal height; 
 \item  find the intersection point $x_d$ of $z=a_r(x)$ and \mbox{$z=\ell_\infty$;}
 \item  define the contact angle of the \emph{d}roplet as 
        \beq \label{ca_planar}
	  \theta_d=\arctan a_r'(x_d) = \arctan \frac{x_d}{\sqrt{R^2-x_d^2}} \, . 
	\eeq
\end{enumerate}

\begin{figure}[htb]
 \begin{center}
     \includegraphics{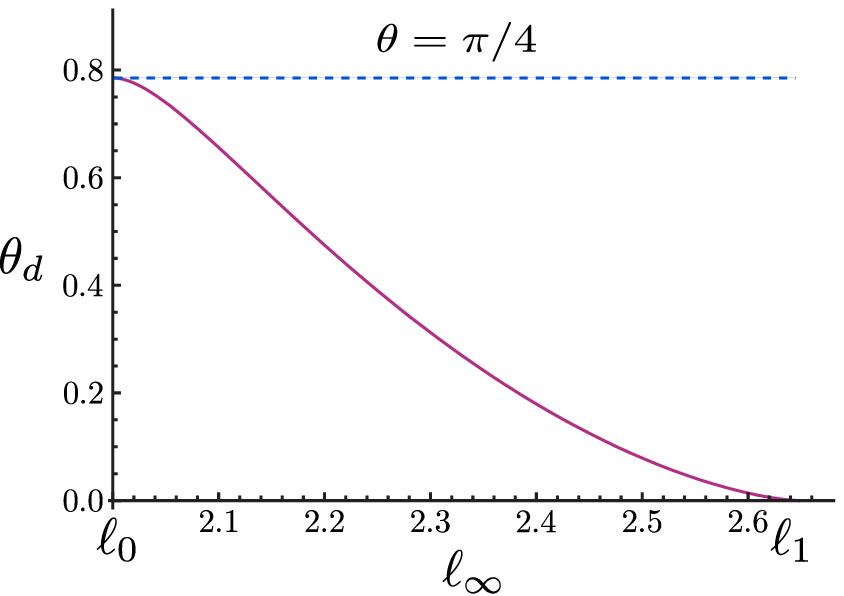} 
  \caption{Dependence of the contact angle $\theta_d$ of the liquid droplet on the thickness at infinity $\ell_\infty$ for the same potential parameters as the ones used in Fig.\,\ref{fig_f0fL}, in particular \mbox{$\ell_0=2$} and \mbox{$\ell_1=2.65$}. The macroscopic limit \mbox{$\theta=\pi/4$} is marked by the horizontal dashed line, and \mbox{$\theta_d(\ell_\infty \to \ell_0)-\theta \sim - (\ell_\infty-\ell_0)^2$}. The limiting value \mbox{$\theta_d=0$} is approached as 
  \mbox{$\theta_d(\ell_\infty \to \ell_1) \sim (\ell_1-\ell_\infty)^{3/2}$}. All lengths are measured in units of $\sigma$. There is a one-to-one correspondence between $\ell_\infty$ and $V_{ex}$ 
   (see Fig.\,\ref{fig_f0fL}).
 \label{fig_thetad}}
 \end{center}
\end{figure}

One can show, that the cosine of the contact angle of the nanodroplet as defined above is given by
\beq
 \cos \theta_d &=& 1 - 
   \frac{\omega_p(f_0)-\omega_p(\ell_\infty)}{\gamma}\left[1-\frac{\omega_p'(f_0)}{\omega_p'(\ell_\infty)} \right] \, .
\eeq
With decreasing thickness $\ell_\infty$ at infinity (i.e., increasing excess volume of the droplet, Fig.\,\ref{fig_f0fL}) the height 
$f_0$ at the center and also the contact angle increases \cite{Weijs2011}. According to Fig.\,\ref{fig_f0fL}, for $V_{ex} \to \infty$ one has 
$\ell_\infty \to \ell_0$ and $f_0 \to \infty$ so that $\omega_p(f_0)$ and $\omega_p'(f_0)$ vanish. Thus, as expected, for large droplets 
the contact angle $\theta_d$ reaches Young's angle $\theta$ (see Eq.\,(\ref{cosine}) and Fig.\,\ref{fig_thetad}). This size dependence of 
$\theta_d$ must be taken into account while investigating Gibbs' criterion for a sessile nanodroplet deposited on a trapezoidal substrate (see below).

\subsection{Trapezoidal substrate}
\subsubsection{Macroscopic description \label{sec_macroscopic}}
\begin{figure}[htb]
 \begin{center}
     \includegraphics{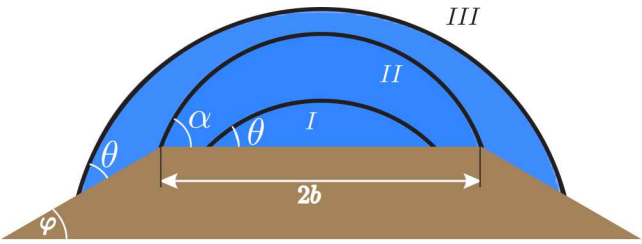} 
    \caption{Three possible types of cylindrical cap-like configurations $I$, $II$, and $III$ of a sessile liquid ridge on a trapezoidal substrate characterized by the angle $\varphi$ and the size $2b$ of the planar basis. The system is translationally invariant in the direction normal to the figure shown. \label{fig_drops}}
\end{center}
\end{figure}

In this section we analyze the shape of a droplet of a fixed volume which is deposited on a trapezoidal substrate characterized by the angle $\varphi$ and by the width $2b$ of the planar basis (Fig.\,\ref{fig_drops}). The system is taken to be translationally invariant in the $y$-direction. The surface free energy $\mathscr{F}$ of the droplet has the following form:
\beq \label{fenergy}
 \mathscr{F} &=& A \gamma + A_{sl} \gamma_{sl} + A_{sg} \gamma_{sg} \, ,
\eeq
where $A$,  $A_{sl}$, and $A_{sg}$ denote the areas of the liquid-gas, solid-liquid, and solid-gas interfaces with the corresponding surface tension coefficients $\gamma$, $\gamma_{sl}$, and $\gamma_{sg}$. Within this macroscopic level of description the effective interaction between the substrate-liquid and the liquid-gas interfaces is not taken into account. We focus on the case that the droplet is deposited  symmetrically on the substrate. Moreover, we restrict our analysis to situations in which Young's local contact angles $\theta$ are restricted to $\theta < \pi/2 - \varphi$, so that even over the sides of the trapezoid the liquid-gas interface can be described by a single-valued function $f=f(x)$, where the $x$-axis is parallel to the planar basis of the trapezoid. In order to describe liquid-gas interfaces with overhangs another parametrization is needed, e.g. by the arc length of the interface. But then the density functional 
(Eq.\,(\ref{fdep})) has a much less transparent form. 

The equilibrium shape of the liquid-gas interface, which minimizes the free energy in Eq.\,(\ref{fenergy}), forms the cap of a cylindrical ridge. Depending on the volume $V$ of the liquid drop one of three distinct types of configurations occurs (Fig.\,\ref{fig_drops}): $(I)$ the area of the substrate-liquid interface is smaller than the area of the horizontal basis of the substrate and the apparent contact angle is equal to Young's angle $\theta$ formed with a horizontal, planar surface; $(II)$ the area of the substrate-liquid interface coincides with that of the horizontal basis with the three-phase contact line pinned to the edge and with the apparent contact angle $\alpha$ formed with the horizontal basis in the range $\theta \leqslant \alpha \leqslant \theta+\varphi$; $(III)$ the area of the substrate-liquid interface exceeds the one of the horizontal basis and the apparent contact angle formed with the tilted side of the trapezoid is again Young's angle. 

For configurations $I$ and $III$ the constant radius $R$ of curvature of the equilibrium liquid-gas interface and its maximal height $h$ above the horizontal basis are given by
\beq
 R_I^2(V) &=& \frac{V/L_y}{\theta -\sin \theta \, \cos \theta }
\eeq
and
\beq
 h_I(V) &=& R_I(V) (1-\cos \theta ) \, , 
\eeq
and by
\begin{align} \label{RIII}
 R_{III}^2(V) \,=\, \frac{V/L_y-b^2\tan \varphi }{\theta \!+\!\varphi\!-\!\sin (\theta \!+\!\varphi ) \cos (\theta \!+\!\varphi )\!-\! 
   \sin ^2(\theta \!+\!\varphi )\tan \varphi } 
\end{align}
and
\begin{align}
 \begin{split}
 h_{III}(V) \,=\,& R_{III}(V)\Big[1-\cos (\theta +\varphi) -\sin (\theta +\varphi) \tan \varphi \Big] \\ & +b \tan \varphi  \, .
\end{split}
\end{align}
$L_y$ is the spatial extension of the system in the invariant $y$-direction.  

\begin{figure}[htb]
 \begin{center}
   \includegraphics{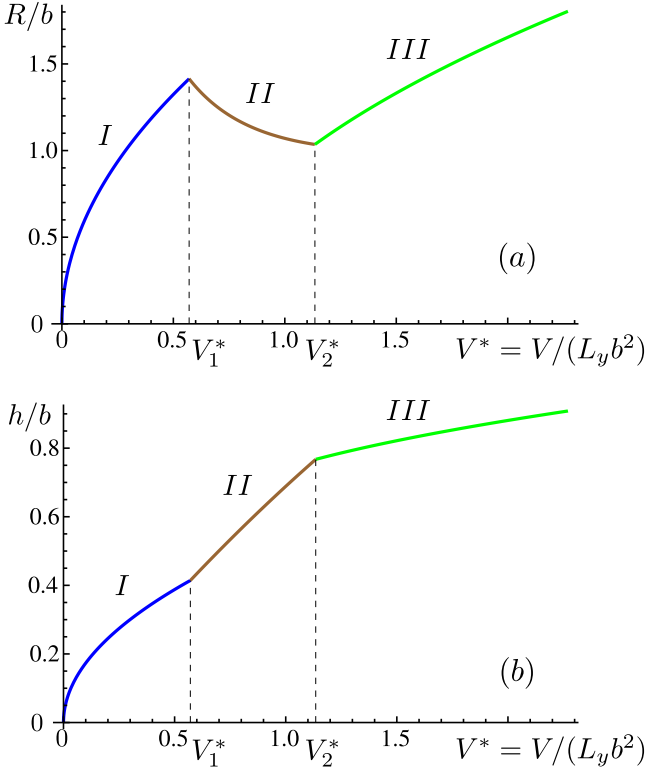}  
    \caption{Dependence of the radius $R$ of curvature $(a)$ and of the maximal height $h$ $(b)$ of the liquid-gas interface shown in 
Fig.\,\ref{fig_drops} on the reduced volume $V^*=V/(L_y b^2)$ of the droplet for $\theta>\varphi$. The length $b$ is half of the width 
of the basis of the trapezoid. $V_1$ and $V_2$ are the limiting values corresponding to configuration $II$ for which the three-phase 
contact line is pinned to the edge of the substrate so that the apparent contact angle $\alpha$ varies in the range 
$\theta \leqslant \alpha \leqslant \theta+\varphi$ (see Fig.\,\ref{fig_drops}), depending on the volume $V$. 
$V_1^*=V(\alpha=\theta)/(L_y b^2)$ and $V_2^*=V(\alpha=\theta+\varphi)/(L_y b^2)$ (see Eq.\,(\ref{mac_alpha})). In 
$(b)$, $h/b$ increases for $I$ and $II$ whereas the curve for $III$ increases for $\theta>\varphi$, is constant for 
$\theta=\varphi$, and decreases for $\theta<\varphi$. In both $(a)$ and $(b)$ the angles $\theta$ and $\varphi$ are chosen as 
$\theta=\pi/4$ and $\varphi=\pi/6$. \label{fig_Randf}}
\end{center}
\end{figure}

In configuration $II$ the apparent contact angle $\alpha$ is not fixed by materials properties. It is not given by Young's equation, but depends on the volume of the sessile droplet and is determined implicitly by the equation
\beq \label{mac_alpha}
 V/(L_y \, b^2) &=& \frac{\alpha}{\sin^2 \alpha} -  \cot \alpha \, .
\eeq
Equation (\ref{mac_alpha}) states that the corresponding section of a circle has the area $V/L_y$. In this case the constant radius $R$ of curvature of the interface and its maximal height are given by
\beq
R_{II}(V) &=& \frac{b}{\sin \alpha} 
\eeq
and
\beq
h_{II}(V) &=& R_{II}(V) (1-\cos \alpha ) \, ,
\eeq
respectively. The dependence of the radius $R$ and of the maximal height $h$ of the interface on the volume are shown in 
Fig.\,\ref{fig_Randf}. The volumes $V_1$ and $V_2$ are the limiting values for configuration $II$, and can be calculated by replacing in Eq.\,(\ref{mac_alpha}) $\alpha$ by $\theta$ and $\theta+\varphi$, respectively. We emphasize that the radius $R$ of the interface is a decreasing function of the volume for configuration $II$ but an increasing function  otherwise, regardless of the angles $\theta$ and $\varphi$. On the other hand the height $h$ of the droplet is an increasing function of the volume in configurations $I$ and $II$. 

For configuration $III$ the height increases with volume for $\theta>\varphi$ (which is compatible with the constraint $\theta<\pi/2-\varphi$ for $\varphi<\pi/4$) and it decreases with volume for $\theta<\varphi$. For $\theta=\varphi$ the height remains constant, i.e., it is volume independent. For fixed angle $\varphi$ there is a minimal value $\theta_{\min}(\varphi)<\varphi$ below which there is no one-drop solution in configuration $III$. The angle $\theta_{\min}(\varphi)$ is the zero of the denominator on the rhs of Eq.\,(\ref{RIII}). For $\theta_{\min}(\varphi) \leqslant \theta \leqslant \varphi$ the volume of the droplet in configuration $III$ is bounded from above by a maximal volume $V^m(\theta,\varphi)$, so that for such a volume the liquid-gas interface touches the edges of the substrate and the droplet splits into three parts. The radius $R^{m}$ of curvature corresponding to $V=V^{m}(\theta,\varphi)$ is given by
\beq
   R_{III}^{m} = \frac{1}{\sin (\varphi-\theta)} \, .  
\eeq
The analysis of morphological phase transitions between distinct sessile droplet configurations on a trapezoidal substrate, at which one 
droplet splits into two or three droplets, is left for future research. This has been already investigated for droplets deposited on 
axisymmetric pillar-like substrates \cite{Du2010,Mayama2011}.

\subsubsection{Mesoscopic description \label{sec_meso}}
In the mesoscopic description one takes into account the presence of the wetting film the droplet is connected with and the effective interface 
potential between the substrate-liquid and the liquid-gas interfaces. The disjoining pressure for the trapezoidal substrate (Fig.\,\ref{fig_drops}) 
is the difference of the disjoining pressures $\Pi_{ap}(x,z,\varphi)$ (Eq.\,(\ref{PI})) corresponding to two apex-shaped substrates (Fig.\,\ref{param_d})
\begin{align}
 S_1 =& \Big\{(x,y,z)\in \mathbb{R}^3: z< (x+b) \tan \varphi  \land z<0  \Big\} \, , \\
 S_2 =& \Big\{(x,y,z)\in \mathbb{R}^3: z<0 \land z>- (x-b)\tan \varphi \Big\} \, , 
\end{align}
for which the characteristic angles are given by $\varphi_1 = \varphi$ and $\varphi_2 = \pi-\varphi$, respectively, with $x=0$ as the center of 
the trapezoidal basis (see Sec.\,\ref{sec_apex}). Thus the disjoining pressure stemming from the trapezoidal substrate is 
\begin{align} \label{disj_pressure}
 \Pi_{trap}(x,z;b,\varphi) &=& \Pi_{ap}(x_1,z_1,\varphi_1) - \Pi_{ap}(x_2,z_2,\varphi_2) 
\end{align}
where
\beq
\binom{x_1}{z_1} = \binom{\quad \cos \frac{\varphi_1}{2} \quad  \sin \frac{\varphi_1}{2}}{ -\sin \frac{\varphi_1}{2} \quad  \cos \frac{\varphi_1}{2}}
    \binom{x+b}{z} 
\eeq
and
\beq 
\binom{x_2}{z_2} = \binom{\quad \cos \frac{\varphi_2}{2} \quad  \sin \frac{\varphi_2}{2}}{-\sin \frac{\varphi_2}{2} \quad  \cos \frac{\varphi_2}{2}}
     \binom{x-b}{z} 
\eeq
are the coordinates corresponding to the above apex-shaped substrates $S_1$ and $S_2$, respectively. 

\begin{figure}[htb]
 \begin{center}
 \includegraphics{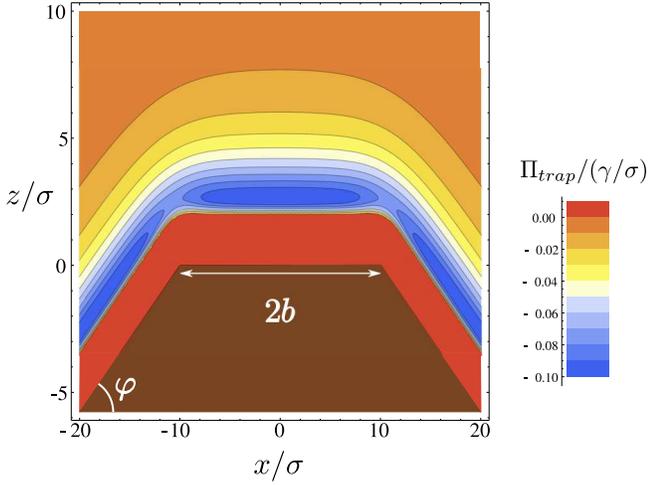}  
 \caption{Disjoining pressure for a trapezoidal substrate with a basis of width $2b=20\sigma$ and a characteristic angle \mbox{$\varphi=\pi/6$}. 
Positive values of the disjoining pressure are summarily marked with red color. The fluid-fluid and the substrate-fluid pair potentials are 
of the van der Waals type with parameters rendering a thickness $\ell_0=2 \sigma$ of the wetting film on a planar substrate and a contact 
angle $\theta=\pi/4. $\label{fig_dptrap}}
 \end{center}
\end{figure}

For the attractive parts of the fluid-fluid and substrate-fluid pair potentials of the van der Waals type (Eq.\,(\ref{VdWaals})), 
the disjoining pressure is positive near the substrate, has two saddle points, and approaches zero from below for points far away 
from the substrate (Fig.\,\ref{fig_dptrap}). 


\subsubsection{Equilibrium shape of the sessile droplet}

\begin{figure}[htb]
 \begin{center}
\includegraphics{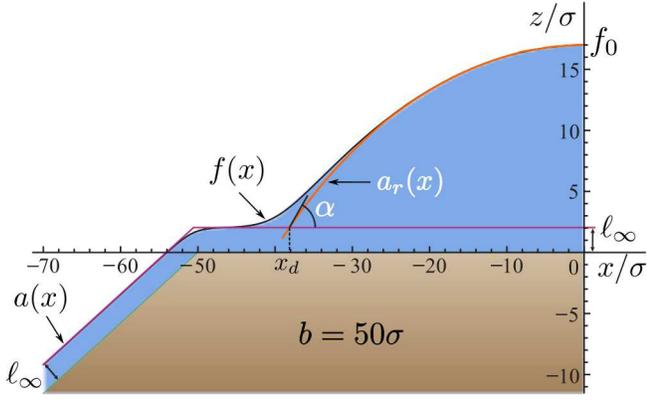}  
 \caption{Equilibrium shape $f(x)$ of the liquid-gas interface for $f_0=17$, which renders $\ell_\infty = 2.04$, and the fitted arc of a circle $a_r(x)$ 
for a sessile droplet deposited on a trapezoidal substrate with a characteristic angle $\varphi=\pi/6$ and a basis width $2b=100\sigma$. 
The thickness of the wetting film far away from the edge is denoted by $\ell_\infty$. The contact angle $\alpha$ is defined as the slope of 
$a_r(x)$ at the intersection $x=x_d$ of $a_r(x)$ with the asymptote $a(x)$ (Eq.\,(\ref{aodx})). The parameters of the effective interface potential are 
$\hat \sigma_s=0.5$, $\theta=\pi/4$, and  $\ell_0 = 2 \sigma$.   \label{fig_wykef}}
 \end{center}
\end{figure}

The effective Hamiltonian for the interface of a sessile droplet deposited on a trapezoidal substrate is given by 
\begin{align} \label{ham_trap}
 \begin{split}
\mathscr{H}_{trap}[f] \!=\! L_y \int_{-\infty}^{\infty} \! \dd x \Big\{ & \gamma \Big[\sqrt{1\!+\!(f'(x))^2} \!-\!\sqrt{1\!+\!((a'(x))^2} \Big] \\
        &  +\omega_{trap}(x,f(x))\!-\! \omega_{trap}(x,a(x))\Big\}\, .
 \end{split}
\end{align}
As the vertical distance from the planar base of the trapezoid the function $f=f(x)$ describes the shape of the liquid-gas interface and 
\begin{align} \label{aodx}
 \begin{split}
 a(x) \,=\, & \ell_\infty \Theta \Big( (b+\ell_\infty \tan \frac{\varphi}{2}) -|x| \Big) \\
  & \!+\!\Big[(b\!-\!|x|) \tan \varphi\!+\!\frac{\ell_\infty}{\cos \varphi} \Big] \Theta \Big(|x|\!-\!(b\!+\!\ell_\infty \tan \frac{\varphi}{2}) \Big)
\end{split}
\end{align}
describes a continuous reference configuration (see the violet line in Fig.\,\ref{fig_wykef}) which contains an a priori unknown thickness 
$\ell_\infty$  as a parameter. The excess volume
\beq \label{Vex}
V_{ex} &=& L_y \int_{-\infty}^{\infty} \dd x \, \Big[\bar f(x) - a(x) \Big] \, 
\eeq
is fixed. The equilibrium profile $f=\bar f(x)$  satisfies the equation (see Eq.\,(\ref{disj_pressure}))
\begin{align} \label{diff_shape}
 \begin{split}
   \gamma \frac{\bar f''(x)}{(1+\bar f'(x)^2)^{3/2}} \,=\,& \left. \frac{\partial \omega_{trap}(x,z)}{\partial z}\right|_{z=\bar f(x)} -\lambda \\
        & = - \Pi_{trap}(x,\bar f(x)) -\lambda \, ,
\end{split}
\end{align}
where $\lambda$ is a Lagrange multiplier. In the following we omit the overbar indicating the equilibrium profile. 

The equilibrium shape of the interface is taken to be symmetric with respect to the plane $x=0$ together with $f'(x=0)=0$. 
Moreover we assume that for $x \to \pm \infty$ the shape $f(x)$ approaches the function $a(x)$. For $x \to \pm \infty$ the 
effective interface potential converges to its planar substrate form 
\begin{align}
 \lim_{x \to \pm \infty} \omega_{trap}(x,f(x)) = \omega_p \Big( \cos \varphi (f(x)\!-\!a(x))\!+\!\ell_\infty \Big) \, ,
\end{align}
so that the Lagrange multiplier $\lambda$ can be determined from the boundary conditions at infinity. Since \mbox{$f''(x\to \pm \infty)=0$} and 
$\Pi_p(\ell)=-\omega'_p(\ell)$, Eq.\,(\ref{diff_shape}) leads to 
\beq \label{lambda_drop}
 \lambda &=& \omega_p'(\ell_\infty) \, .
\eeq

For a given value of $\ell_\infty$ the solution $f(x)$ of the second order differential equation (\ref{diff_shape}) contains no free parameter. 
The two integration constants are determined by the boundary conditions \mbox{$f'(x=0)=0$} and \mbox{$f(x \to -\infty) = (x+b) \tan \varphi+\ell_\infty/\cos \varphi$}. 
On the other hand the parameter $\ell_\infty$ is determined by the excess volume $V_{ex}$ of the droplet (Eq.\,(\ref{Vex})).

In order to find the equilibrium profile of the liquid-gas interface we use a procedure analogous to the one used in Sec.\,\ref{sec_appprofile}:
\begin{enumerate}
 \item fix $f(x=0)$ at a certain value $f_0>0$;
 \item fix $\ell_\infty>\ell_0$ and calculate the Lagrange multiplier $\lambda$ (Eq.\,(\ref{lambda_drop}));
 \item integrate Eq.\,(\ref{diff_shape}) numerically with the boundary conditions $f(0)=f_0$ and $f'(0)=0$ within the range $x \in [-L,0]$, where 
$x=-L$ is the imposed limit of the system size on the left  hand side;
 \item compare $f(-L)$ with $a(-L)$ and $f'(-L)$ with $a'(-L) = \tan \varphi$;
 \item if the differences $|f(-L)/a(-L)-1|$ and $|f'(-L)/a'(-L)-1|$ are not satisfyingly small return to step 2 and use a different choice for 
$\ell_\infty$.
\end{enumerate}
This procedure fixes $f(x=0)$, and $\ell_\infty$ and thus $V_{ex}$ follow; this relationship can be inverted. Our method is restricted to contact 
angles $\theta$ within the range $[\varphi, \pi/2 - \varphi]$. At the upper limit $\pi/2-\varphi$ the liquid-gas interface can develop overhangs 
which cannot be described by a single-valued function $f=f(x)$. On the other hand, for $\theta \leqslant \varphi$ there are many $\ell_\infty$ 
corresponding to the same $f_0$ and it is not obvious how to choose the new value of $\ell_\infty$ when returning from step 5 to step 2 in the 
above algorithm. This problem becomes already apparent within the macroscopic description according to which in the case $\theta<\varphi$
the height $h$ of the droplet is the same for two distinct volumes (see $III$ in Fig.\,\ref{fig_Randf}\,(b)).

\begin{figure}[htb]
 \begin{center}
 \includegraphics{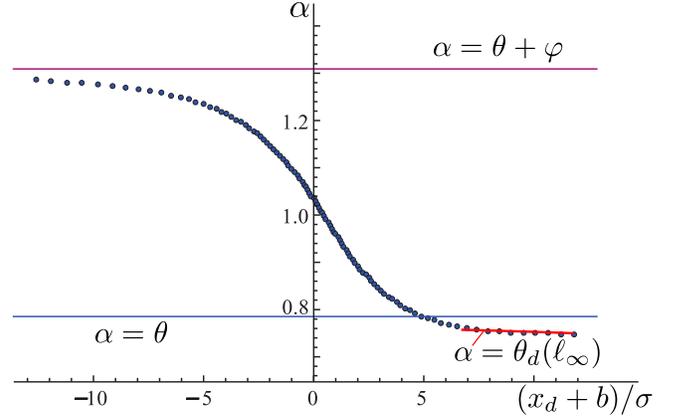}  
 \caption{Dependence of the contact angle $\alpha$ on the shifted position $x_d+b$ of the three-phase contact line for the equilibrium shape of the 
sessile nanodroplets as the one shown in Fig.\,\ref{fig_wykef}. The horizontal lines denote the angles $\theta$ and $\theta+\varphi$ expected from 
Gibbs' criterion. The short line $\alpha=\theta_d(\ell_\infty)$ on the right end corresponds to the contact angles for nanodroplets deposited on 
the same but planar substrate for the same values of $\ell_\infty$ as the ones rendering the data points there, for which there is a one-to-one 
correspondence between $x_d$ and $\ell_\infty$ as well as between $\ell_\infty$ and $V_{ex}$ (see for comparison the lower curve in Fig.\,\ref{fig_f0fL} 
corresponding to a planar substrate). The parameters of the effective interface potential and the angle $\varphi$ are the same as in Fig.\,\ref{fig_wykef}
and $b=50 \sigma$.   \label{fig_alfa}}
\end{center}
\end{figure}

For each solution of the equilibrium shape of the liquid-gas interface we fit the arc of a circle \mbox{$a_r(x)=\sqrt{R^2-x^2}-R+f_0$} with the same 
curvature as the one of the liquid-gas interface at $x=0$ (Fig.\,\ref{fig_wykef}). The radius of the arc of the circle is given by 
(see Eqs.\,(\ref{diff_shape}) and (\ref{lambda_drop}))
\beq
 R = \frac{\gamma}{\Pi_{trap}(0,f_0)+\omega_p'(\ell_\infty)} \, .
\eeq
In addition, for each equilibrium profile the position $x=x_d<0$ at which the arc of the circle intersects the asymptote function, $a_r(x_d)=a(x_d)$, 
and the corresponding angle $\alpha = \arctan a_r'(x_d)$ are determined. Upon increasing $f_0$ (i.e., increasing the volume of the droplet) the position 
$x_d$ moves smoothly across the edge of the trapezoidal substrate and the angle $\alpha$ increases (Fig.\,\ref{fig_alfa}). The latter is bounded from 
above by $\theta+\varphi$ in accordance with Gibbs' criterion. However, for $x_d$ sufficiently far to the right of the edge one finds $\alpha<\theta$ 
(Fig.\,\ref{fig_alfa}) which is in contradiction with Gibbs' criterion for macroscopic droplets. 

For the equilibrium solutions characterized by $f_0$ (or, equivalently, $\ell_\infty$ or $V_{ex}$) we have calculated the contact angle 
$\theta_d(\ell_\infty)$ of the droplet deposited on a planar substrate (Eq.\,(\ref{ca_planar})). It turns out that the function 
$\alpha(x_d)$ tends to $\theta_d(\ell_\infty)$ for $x_d+b>0$ (Fig.\,\ref{fig_alfa}). This behavior can be understood by noting 
that for nanodroplets the contact angle $\theta_d$ changes significantly with their volume (Fig.\,\ref{fig_thetad}). The application 
of Gibbs' criterion also to nanodroplets states that the contact angle of the sessile droplet near the edge is bounded from below by 
the contact angle of the corresponding droplet deposited on the same but planar substrate and from above by the contact angle 
$\theta+\varphi$, as for macroscopic droplets. 

We recall that the contact angle of nanodroplets depends sensitively on their volume. For macroscopic droplets $\theta_d \to \theta$ 
and in Fig.\,\ref{fig_alfa} the difference between Young's angle $\theta$ and $\theta_d(\ell_\infty)$ would vanish. In this limit the 
shape of the function $\alpha(x_d+b)$ would resemble the one obtained for the liquid-gas interface adsorbed at an apex-shaped 
substrate (Fig.\,\ref{crit}). Macroscopic droplets deposited on a trapezoidal substrate with $x_d+b>0$ can be prepared for macroscopic 
values of the width $2b$.

\begin{figure}[htb]
 \begin{center}
 \includegraphics{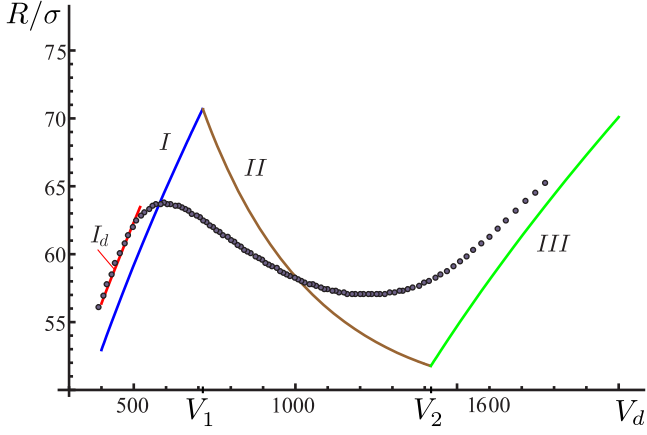}  
 \caption{Dependence of the radius of curvature $R$ of liquid droplets deposited on a trapezoidal substrate (Fig.\,\ref{fig_wykef}) 
on their volume $V_d$ within the mesoscopic (dots) and the macroscopic (full lines, Fig.\,\ref{fig_Randf}$(a)$) description. The line 
$I_d$ denotes the radius calculated within the macroscopic description but taking into account the change with volume of the contact 
angle $\theta_d$ of the corresponding nanodroplet on the planar substrate. The surface tension coefficient, the parameters of the 
effective interface potential, and the angle $\varphi$ are the same as in Fig.\,\ref{fig_wykef}. The droplet volume is measured 
in units of $L_y\sigma^2$. \label{fig_R_both}}
\end{center}
\end{figure}

\begin{figure}[htb]
 \begin{center}
 \includegraphics{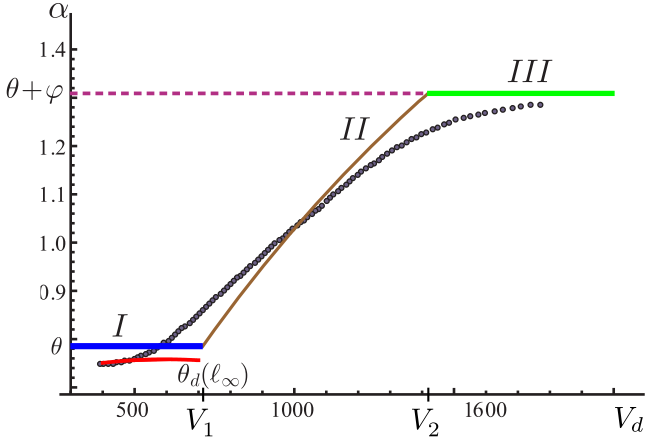} 
 \caption{Dependence of the contact angle $\alpha$ of the liquid droplet deposited on a trapezoidal substrate on the droplet volume 
$V_d$ within the mesoscopic (dots) and the macroscopic (full line) description (compare Fig.\,\ref{fig_Randf}). The short line 
$\theta_d(\ell_\infty)$ denotes the contact angle calculated for the droplets deposited on the same but planar substrate with 
the values of  $\ell_\infty$ corresponding to the ones for the dotted line. (According to Fig.\,\ref{fig_alfa} there is a one-to-one correspondence between $\ell_\infty$ and $V_d \simeq V_{ex}$.) The parameters of the underlying effective interface 
potential and the angle $\varphi$ are the same as in Fig.\,\ref{fig_wykef}. Figure \ref{fig_V_both} translates Fig.\,\ref{fig_alfa} 
into the dependence on the droplet volume, which is measured in units of $L_y\sigma^2$. \label{fig_V_both}}
\end{center}
\end{figure}

\begin{figure}[htb]
 \begin{center}
 \includegraphics{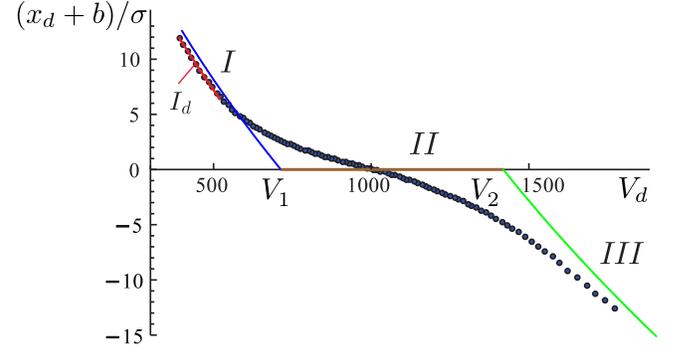} 
 \caption{Dependence of the distance $x_d+b$ from the edge of the substrate of the three-phase contact line (Fig.\,\ref{fig_wykef}) of 
liquid droplets deposited on a trapezoidal substrate on the droplet volume $V_d$ within the mesoscopic (dots) and the macroscopic 
(full lines) description. The full line $II$ describes the macroscopic pinning for $V_1<V<V_2$ (compare Fig.\,\ref{fig_Randf}). 
The short line $I_d$ denotes $x_d+b$ as calculated within the macroscopic description but taking into account the change of the 
contact angle of the nanodroplet $\theta_d$ with volume. The parameters of the underlying effective interface potential and the angle 
$\varphi$ are the same as in Fig.\,\ref{fig_wykef}. The droplet volume is measured in units of $L_y\sigma^2$. \label{fig_x0_both}}
\end{center}
\end{figure}

For simplicity instead of the excess volume $V_{ex}$ we use the volume $V_d$ defined as the excess volume corresponding to the function 
$a_r(x)$ over the function $a(x)$ in order to characterize the volume of the droplet approximately. Within a mesoscopic description the 
radius of curvature $R$, the contact angle $\alpha$, and the parameter $x_d$ are smooth functions of $V_d$, which is in contrast to the 
macroscopic description (Figs. \ref{fig_R_both} -- \ref{fig_x0_both}). For large volumes these quantities approach their macroscopic 
analogues. It turns out that for small volumes they can be described by the corresponding macroscopic equations if one uses the volume 
dependent contact angle $\theta_d$.  


\begin{figure}[htb]
 \begin{center}
 \includegraphics{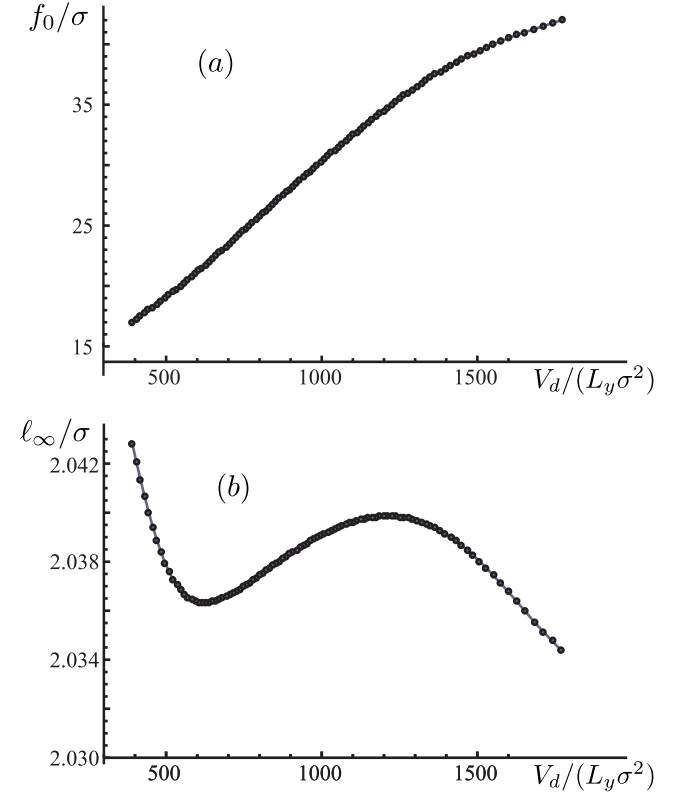} 
 \caption{Dependence of the maximal height $f_0$ $(a)$ and of the wetting film thickness $\ell_\infty$ $(b)$ of liquid droplets deposited 
on a trapezoidal substrate on their volume $V_d$. The parameters of the underlying effective interface potential and the angle $\varphi$ 
are the same as in Fig.\,\ref{fig_wykef}. \label{fig_f0finf}}
\end{center}
\end{figure}

For effective interface potentials rendering $\theta>\varphi$ the height $f_0$ of the droplet is an increasing function of its volume 
$V_d$ (Fig.\,\ref{fig_f0finf}$(a)$) as in the macroscopic case (Fig.\,\ref{fig_Randf}$(b)$). The thickness $\ell_\infty$ (Fig.\,\ref{fig_f0finf}$(b)$) is a non-monotonic 
function of the volume (in contrast to the planar case (see Fig.\,\ref{fig_f0fL})), which signals the transition from configurations 
$I$ to $III$ introduced for macroscopic droplets (Sec.\,\ref{sec_macroscopic}). 

\cleardoublepage 

\subsubsection{Width of the transition region}
Within the macroscopic description the three-phase contact line remains pinned at the edge of the trapezoidal substrate for a certain range 
$V_1<V<V_2$ of droplet volumes (see Fig.\,\ref{fig_x0_both}). For such configurations (denoted as $II$ in Sec.\,\ref{sec_macroscopic}) 
the radius of the droplet decreases with its volume (Fig.\,\ref{fig_Randf}$(a)$) while for configurations $I$ and $III$ it is an increasing 
function of the volume. In the mesoscopic description there is no contact line pinning. However, the radius $R$ of the arc of the circle  fitted 
to the equilibrium liquid-gas interface has a similar non-monotonic volume dependence as in the macroscopic description (Fig.\,\ref{fig_R_both}).
\begin{figure}[htb]
 \begin{center}
 \includegraphics{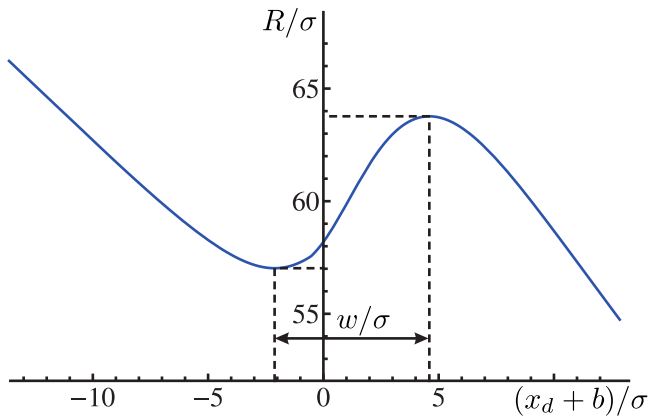}  
 \caption{Dependence of the radius $R$ of the arc of the circle fitted to the interface shape on the distance $x_d+b$ of the three-phase 
contact line from the edge of the substrate (see Fig.\,\ref{fig_wykef}). The distance $w$ between the abscissa of the local extrema is a 
measure of the width of the transition region. The parameters of the underlying effective interface potential and the angle $\varphi$ are 
the same as in Fig.\,\ref{fig_wykef}. \label{fig_R2}}
 \end{center}
\end{figure}
 According to Figs.\,\ref{fig_R_both} and \ref{fig_x0_both} the region where $R$ is an increasing function of $x_d+b$ corresponds to the 
aforementioned pinning in the macroscopic description. The spatial extent of this region is denoted by $w$ and is a measure of the width 
of the transition region within which the contact line passes smoothly across the edge (Fig.\,\ref{fig_R2}). For the choice of the 
parameters used in Fig.\,\ref{fig_R2} the width of the transition region is of the order of ten fluid particle diameters and thus 
is mesoscopic in character. 


\subsubsection{Line contribution to the free energy}
For the system under investigation we define the line contribution to the free energy as the difference between the free energy of the 
droplet and the free energy of the configuration corresponding to the arc of the circle $a_r(x)$ fitted to the droplet, both relative 
to the reference configuration $a(x)$:
\begin{align}
 \begin{split}
 \tau[f] = &  (\mathscr{H}_{trap}[f]- \mathscr{H}_{trap}[a_r])/L_y = \\
   & 2\int_{-\infty}^0 \dd x \, \Big\{ \gamma \Big[\sqrt{1+(f'(x))^2}-\sqrt{1+(a'(x))^2} \Big] \\ & +\omega_{trap}(x,f(x))-\omega_{trap}(x,a(x))\Big\} \\
   &   -2 \int_{x_d}^{0} \dd x \, \Big\{ \gamma \Big[\sqrt{1+(a_r'(x))^2}-\sqrt{1+(a'(x))^2} \Big] \\ & +\omega_{trap}(x,a_r(x))-\omega_{trap}(x,a(x))\Big\} 
\, .
 \end{split}
\end{align}

\begin{figure}[htb]
 \begin{center}
 \includegraphics{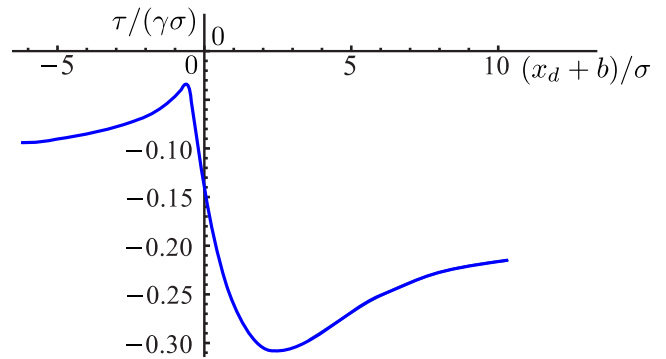}    
 \caption{Dependence of the line contribution $\tau$ on the distance $x_0+b$ of the three-phase contact line from the edge of the substrate of a sessile 
droplet deposited on a trapezoidal substrate. The parameters of the underlying effective interface potential and the angle $\varphi$ are the same as on 
Fig.\,\ref{fig_wykef}.  \label{fig_wom2}}
\end{center}
\end{figure}

The line contribution changes significantly when the three-phase contact line passes the edge (Fig.\,\ref{fig_wom2}). The spatial extent of the region in 
which the line contribution is an increasing function of the volume of the droplet (i.e., decreasing as function of $x_d+b$) is of the order of three 
fluid particle diameters. If the contact line is far from the edge the line contribution is a decreasing function of the volume of the droplet. 

We emphasize that the line contribution to the free energy presented in Fig.\,\ref{fig_wom2} corresponds to different equilibrium profiles of the 
droplets, in particular with different volumes. It is not the plot of the line energy of a droplet with a fixed excess volume. Nonetheless, 
Fig.\,\ref{fig_wom2} indicates that there is a free energy barrier at the edge of the substrate and thus a moving droplet with fixed excess 
volume is expected to stop just before reaching it \cite{Moosavi2006,Moosavi2009}.


\section{Summary and discussion}
 \subsection{Summary}

If a substrate surface forms a sharp corner and the three-phase solid-liquid-gas contact line of a sessile droplet is pinned at the substrate apex 
the modified Young's equation for the contact angle (Eq.\,(\ref{mYoung})) is no longer valid. Instead, the corresponding local contact angle  
$\alpha$ can take any value within the range $\theta \leqslant \alpha \leqslant \theta+\varphi$ (Eq.\,(\ref{gibbs_criterion})), where 
$\pi-\varphi$ is the opening angle of the apex formed by the substrate faces (Fig.\,\ref{fig_gibbs}). This ambiguity of the local 
contact angle at the apex is called Gibbs' criterion. In order to determine the equilibrium shape of the liquid-gas interface of a liquid 
film covering an apex-shaped substrate (Fig.\,\ref{fig_apex}) and the equilibrium shape of a sessile droplet deposited on a trapezoidal 
substrate (Fig.\,\ref{fig_drops}) we have used an effective interface Hamiltonian based on density functional theory. This approach has 
proved to be very useful in analyzing similar systems \cite{Hofmann2010}. The thermodynamic state of the system is taken to be at the bulk 
liquid-gas coexistence line below the wetting temperature and well below the critical point of the liquid. For our explicit calculations 
we have chosen the thickness of the wetting film to be of the order of a few fluid particle diameters. We have focused on quasi 
two-dimensional systems being translationally invariant in one direction. 

First, we have analyzed the equilibrium shape $\bar f(x)$ of the liquid-gas interface at a planar substrate (Fig.\,\ref{schem_flat}) 
with boundary conditions \mbox{$\bar f(x\to -\infty) = \ell_0$} and \mbox{$\bar f'(x\to \infty) = \tan \theta$}. The film thickness $\ell_0$ minimizes the 
effective interface potential for a planar substrate $\omega_p(\ell)$ and the angle $\theta$ fulfills the macroscopic Young's 
law (Eqs.\,(\ref{Youngslaw}) and (\ref{Youngslaw2})). If the height of the interface at one point $x_0$ is fixed as \mbox{$f_0=f(x_0)$}, 
the derivative \mbox{$f'_0=f'(x_0)$} is a unique function of $f_0$ (Fig.\,\ref{fprime}) for the aformentioned boundary conditions 
\mbox{$\bar f(x\to -\infty) = \ell_0$} and \mbox{$\bar f'(x\to \infty) = \tan \theta$}. 
It changes monotonously from \mbox{$f'_0=0$} for \mbox{$f_0=\ell_0$} to \mbox{$f'_0=\tan \theta$} for \mbox{$f_0 \to \infty$}. 

In the case of an apex-shaped substrate we have calculated the equilibrium profile for the liquid-gas interface numerically. 
Due to  limited numerical accuracy, for a fixed value $f_0$ one cannot find the value of the derivative $f_0'$ rendering 
the exact boundary condition on the far left hand side $x=L_1$ of the system. What can be achieved numerically is to find values 
$f_0'=f_<'$ and $f_0'=f_>'$ rendering solutions $f_<(x)$ and $f_>(x)$, respectively, which for sufficiently large, negative 
$x$ follow the asymptote $a_{ap}(x)$ (Eq.\,(\ref{eq_asymptote})) and differ only slightly from it in the vicinity of the boundary $x=L_1$ (Fig.\,\ref{mniej}). 
The equilibrium profile $\bar f(x)$ lies between these functions $f_<(x)$ and $f_>(x)$. For each solution $\bar f(x)$ we determine 
the contact angle $\alpha = \arctan \bar f'(L_2)$ (where $L_2$ is the imposed boundary at the right hand side of the system) and 
the quantity $x_d$ characterizing the position at which the liquid-gas interface detaches from the substrate (Fig.\,\ref{param_d}). 
The contact angle $\alpha$ is a decreasing and \emph{continuous} function of $x_d$ (Fig.\,\ref{crit}); there is no indication of three-phase contact line pinning. 
For $x_d \to \pm \infty$ the contact angle $\alpha$ approaches from below its limiting values $\theta \mp \varphi/2$, 
which are those expected from Gibbs' criterion for this geometry. 

In Sec.\,IV we have examined cylindrical droplets with a fixed excess volume as well as incomplete wetting films deposited on planar and trapezoidal substrates. 
In the planar case (Fig.\,\ref{fig_schemdrop}) we have obtained the equation for the shape of the interface for various excess volumes of the 
droplet \cite{Yeh1999a,Yeh1999b,Bertozzi2001,Glasner2003,Gomba2009,Weijs2011}. Both in a finite system of width $2L$ and in the case when the droplets 
can spread over an unbounded substrate, i.e., $L \to \infty$, characteristic features of the droplet shape can be inferred from tangential 
constructions to the effective interface potential $\omega_p(\ell)$ (Figs.\ \ref{fig_constr_sk} and \ref{fig_constr}). This involves in particular the heights $f_L$ (i.e., $\ell_\infty$ in unbounded systems) and $f_0$ denoting the minimal and maximal values, respectively, of the function $\bar f(x)$ describing the shape of the equilibrium nanodroplets. In laterally unbounded systems, $f_0 \to \infty$ and $\ell_\infty \to \ell_0$ for increasing excess volumes $V_{ex} \to \infty$ (Fig.\,\ref{fig_f0fL}). 

For a laterally finite, planar system of size $L$ the droplet solution does not exist for arbitrary 
$f_L > \ell_0$. If $f_L$ is sufficiently small there is no $f_0$ which would correspond to the prescribed fixed $L$ (Fig.\,\ref{fig_Lf0fL}). 
On the other hand the quantity $f_L$ is bounded from above, rendering upper ($V_{tot}^{max}(L)$) and  lower ($V_{tot}^{min}(L)$) bounds for the total volume for which the droplet configuration exists (Fig.\,\ref{fig_VminVmax}).

We have also calculated the contact angle for nanodroplets (Fig.\,\ref{fig_drop}). It is smaller than Young's angle for macroscopic droplets \cite{Weijs2011} and it is an increasing function of the volume of the droplet (Fig.\,\ref{fig_thetad}). This dependence has to be taken into account also for analyzing sessile \emph{nano}droplets deposited on trapezoidal substrates. 

For symmetrical ridgelike macroscopic droplets deposited on trapezoidal substrates one can distinguish three different configurations depending on the position of three phase contact line (Fig.\,\ref{fig_drops}). The radius $R$ of curvature of the droplets is a continuous but neither a smooth nor a monotonous function of the volume of the droplet. It is decreasing if the three-phase conctact line is pinned to the edge of the substrate (configuration $II$), and increasing otherwise (configurations $I$ and $III$) (Fig.\,\ref{fig_Randf}). 

In order to determine the equlibrium shape of nanodroplets on trapezoidal substrates we have calculated the disjoining pressure for such substrates (Fig.\,\ref{fig_dptrap}). For each ensuing equilibrium shape of the liquid-gas interface we have fitted the arc of a circle with the same radius of curvature $R$ as the one of the liquid-gas interface at the center $x=0$ (Fig.\,\ref{fig_wykef}). In addition, the position $x=x_d<0$ at which the arc of the circle $a_r(x)$ intersects the asymptote $a(x)$ of the film thickness and the corresponding contact angle $\alpha = \arctan a_r'(x_d)$ were determined. Upon increasing $f_0$ (i.e., increasing the volume of the droplet) the position $x_d$, which can be interpreted as the three-phase contact line position, moves smoothly across the edge of the trapezoidal substrate and the contact angle $\alpha$ increases (Fig.\,\ref{fig_alfa}). The latter is bounded from above by $\theta+\varphi$ in accordance with Gibbs' criterion. However, for $x_d$ sufficiently far to the right of the edge one finds $\alpha<\theta$ which is in contradiction with Gibbs' criterion for macroscopic droplets. This behavior can be understood by noting that for nanodroplets the contact angle $\theta_d$ changes significantly with their volume (Fig.\,\ref{fig_thetad}). The extension of Gibbs' criterion to nanodroplets states that the contact angle of sessile droplets near the edge of the substrate is bounded from below by the contact angle of the corresponding finite-sized droplets deposited on the same but planar substrate and from above, as for macroscopic droplets, by the contact angle $\theta+\varphi$. 

Within a mesoscopic description the radius of curvature $R$, the contact angle $\alpha$, and the parameter $x_d$ are smooth functions of the volume of the droplet $V_d$, which is in contrast to the macroscopic description (Figs.\,\ref{fig_R_both} -- \ref{fig_x0_both}). For large volumes these quantities approach their macroscopic analogues. It turns out that in the limit of small volumes they can be described by the corresponding macroscopic equations if one uses the volume dependent contact angle $\theta_d$ of nanodroplets.

For effective interface potentials rendering $\theta>\varphi$ the height $f_0$ of the droplet is an increasing function of its volume $V_d$ (Fig.\,\ref{fig_f0finf}), as in the macroscopic case (Fig.\,\ref{fig_Randf}$(b)$). Opposite to the planar case (Fig.\,\ref{fig_f0fL}) the film thickness $\ell_\infty$ is a non-monotonic function of the drop volume, which signals the transition from configuration $I$ to $III$ introduced for macroscopic droplets. According to Fig.\,\ref{fig_R_both} the region where $R$ is a decreasing function of $V_d$ corresponds to three-phase contact line pinning within the macroscopic description (configuration $II$). The spatial extent $w$ of this region is a measure of the width of the transition region within which the contact line passes smoothly across the edge (Fig.\,\ref{fig_R2}). For the choice of the parameters used in Fig.\,\ref{fig_R2} the width of the transition region is of the order of ten fluid particle diameters and thus is mesoscopic in character. 

The line contribution to the free energy of the droplet changes significantly when the three-phase contact line passes the edge (Fig.\,\ref{fig_wom2}). The spatial extent of the region in which the line contribution is an increasing function of the volume of the droplet (i.e., decreasing as function of the distance $x_d+b$ of the three-phase contact line from the edge) is of the order of three fluid particle diameters and thus also mesoscopic in character. If the contact line is far from the edge the line contribution is a decreasing function of the volume of the droplet. The edge poses a free-energy barrier for the three-phase contact line. 

Our numerical results for Gibbs' criterion have been obtained from the analysis of the disjoining pressure for apex-shaped substrates for specific choices of the effective interface potential (based on long-ranged interparticle interactions) and for specific geometrical parameters. We have studied the absence of three-phase contact line pinning on the nanoscale and we have analyzed how Gibbs' criterion has to be modified in order to describe sessile nanodroplets on substrates with sharp edges. 

\subsection{Discussion}
Three-phase contact line pinning which takes place at asperities of non-planar, chemically homogeneous surfaces is a common phenomenon due to inherent 
roughness of both naturally occuring as well as fabricated substrates.  Important examples vary from  capillary filling of geometrically patterned channels to terraced substrates. In the case of channels patterned by pillars, depending on the distance between these obstacles, the width of the channel, and Young's contact angle the advancing liquid front can be pinned and flow can be suppressed \cite{Kusumaatmaja2008,Mognetti2009,Chibbaro2009}. In Sec.\,\ref{sec_apex} we have shown that within a mesoscopic description there is no three-phase contact line pinning of the liquid-gas interface at a substrate edge due to the extra cost related to the associated increase of the line contribution to free energy of the system. Thus one can speculate that dense arrangements of obstacles favor capillary flow in microchannels despite 
the limitations predicted by the macroscopic version of Gibbs' criterion. This conjecture seems to be supported by the fact that dewetting of terraced substrates can proceed for step heights up to a couple of nanometers while under the same thermodynamic conditions dewetting is suppressed for larger step heights \cite{Ondarcuhu2005}.

Three-phase contact line pinning has also an impact on contact angle hysteresis on rough surfaces \cite{Quere2005}. Besides the shape and the chemical 
character of the asperities, their size plays an important role. Already in early studies of three-phase contact line pinning, it was shown that steps 
with a height of approximately $10 \, nm$ do not pin the three-phase contact line at the edge of stepped substrate \cite{Mori1982}. Advances in the 
fabrication of nanostructured substrates have allowed the investigation of the contact line behavior at rings grown on a flat substrate with a 
trapezoidal vertical cross section. For rings with heights below $2\, \mu m$ the advancing contact angle decreases significantly with the height 
of the trapezoidal asperity \cite{Kalinin2009}. Both these observations deviate from the macroscopic Gibbs' criterion and can be related to the 
findings presented in Sec.\,\ref{sec_droplet}, according to which the position of the three-phase contact line moves continuously with a smooth 
variation of the apparent contact angle near the edge of the substrate. More quantitative analyses concerning the suppression of three-phase 
contact line pinning at nanometer sized steps are warranted and promising.

\begin{figure}[htb]
 \begin{center}
 \includegraphics{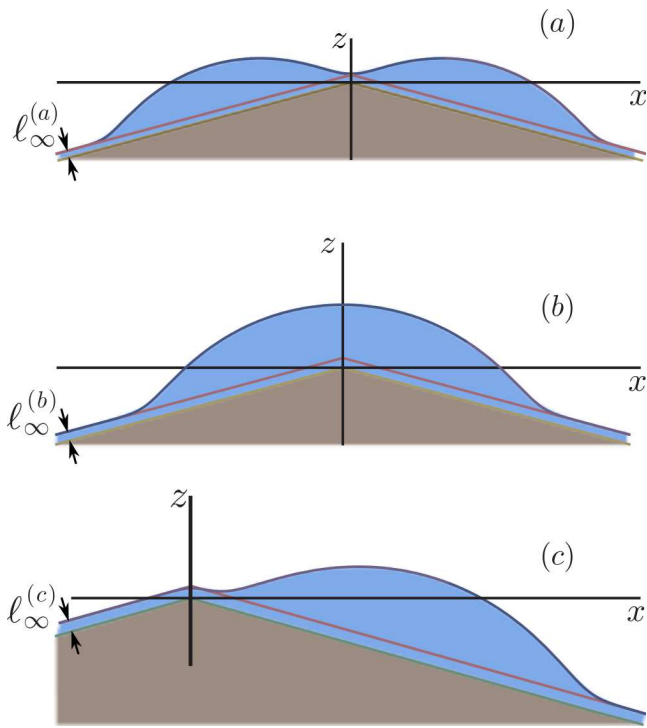}  
 \caption{Schematic shapes of three morphologically distinct equilibrium droplet configurations with the same fixed excess volume deposited on an 
apex-shaped substrate. The configurations $(a)$ and $(b)$ are symmetric with respect to the midplane. The three configurations have different 
asymptotic thicknesses: $\ell_{\infty}^{(a)}$, $\ell_{\infty}^{(b)}$, and $\ell_{\infty}^{(c)}$. \label{fig_morphologies}}
 \end{center}
\end{figure}

For sessile nanodroplets with a fixed excess volume deposited on a laterally unbounded apex-shaped substrate, there are three morphologically 
distinct solutions of the Euler-Lagrange equation (Fig.\,\ref{fig_morphologies}). Two of them are symmetric. For the same excess volume the three 
configurations have different asymptotic thicknesses $\ell_\infty$ and therefore it is not clear a priori which configuration has the lowest free 
energy and thus is the stable one. 

The disjoining pressure for the liquid-gas interface at a trapezoidal substrate can be expressed in terms of the difference of the disjoining 
pressures stemming from two suitable apex-shaped substrates (Eq.\,(\ref{disj_pressure})). Thus also for a trapezoidal substrate there are different 
configurations of the liquid-gas interface fulfilling the Euler-Lagrange equation, as for the apex-shaped substrate case. 
                                                                                                                                                                                                                                                                                                                                                                                                                                                   
Here our investigation has been focused on sessile nanodroplets which are symmetric and attain their maximal  height at the center of the system. 
The issue of morphological transitions between different sessile droplet configurations on apex-shaped and trapezoidal substrates is left for future 
research. This has been already investigated for macroscopic droplets deposited on axisymmetric pillar substrates \cite{Du2010,Mayama2011}. 
Recent studies of the Vapor-Liquid-Solid mechanism of nanowire growth show that the liquid droplet promoting the solid growth can wet the sidewall of the 
nanowire and thus does not sit at the top of the pillar with the three-phase contact line pinned to its edge as for typical VLS growth \cite{Dubrovskii2011}. 
A theoretical description of the transition between these two configurations using the present mesoscopic approach appears to be interesting. \\

Finally, we mention an interesting process in which a droplet of fixed volume $V$ is placed on a trapezoidal substrate and its contact angle $\theta$ is decreased, e.g., by increasing an applied voltage as in electrowetting \cite{Buehrle2003,Mugele2005,Mugele2007}. Initially, the droplet shape corresponds to configuration $I$ on Fig.\,\ref{fig_drops}. Upon decreasing the contact angle the droplet spreads until the three-phase contact line reaches the edge of the trapezoidal substrate. If the corresponding contact angle fulfills \mbox{$\theta=\alpha > \varphi$} (where $\alpha$ is the solution of Eq.\,(\ref{mac_alpha})), upon further increase of the voltage, the shape of the droplet and the apparent contact angle remain constant until the angle $\theta$ decreases to the value $\alpha - \varphi$ (in agreement with Gibbs' criterion).  Upon further increase of the voltage  one expects the contact angle $\theta$ to start to decrease again and the drop to spread  on the tilted side of the trapezoidal substrate. In actual experimental settings the above naive scenario may be substantially modified by effects related to the fact that the drop is charged. In particular the change of morphology of the droplet front upon reaching the apex in the presence of electric fields provides interesting scientific perspectives.

In summary, we have shown that the presence of mesoscopic wetting films on edged substrate surfaces prevents three-phase contact line pinning on the nanoscale. We have analyzed the shape of the liquid-gas interface of liquid films both at an apex-shaped substrate and for liquid sessile nanodroplets with fixed excess volume deposited on trapezoidal substrates. Near the edge of an apex-shaped substrate the apparent contact angle changes continuously within the range of values expected from Gibbs' criterion while the three-phase contact line smoothly passes through the atomically sharp apex. For a trapezoidal substrate, upon increasing the volume of the nanodroplet the apparent contact angle fulfills a modified Gibbs' criterion, for which one has to take into account the dependence of the contact angle of the nanodroplet on its volume. In both cases the spatial extent of the region, within which the three-phase contact line passes across the edge, is of the order of ten fluid particle diameters and thus is mesoscopic in character.

\end{document}